\documentclass[twocolumn]{aastex6}

\usepackage{mathtools}
\usepackage{wasysym}
\usepackage{graphicx}
\usepackage{epstopdf}
\usepackage{xfrac}
\usepackage{textcomp}

\newcommand{\RA}[4]{{#1}$^\text{h}${#2}$^\text{m}${#3}$^\text{s}${#4}}
\newcommand{\dec}[4]{{#1}\textdegree {#2}'{#3}''{#4}}

\shorttitle{Metallicity of dwarf galaxies}
\shortauthors{K. Douglass, M. Vogeley}

\begin{document}

\title{Determining the large-scale environmental dependence of gas-phase metallicity in dwarf galaxies}
\author{Kelly A. Douglass, Michael S. Vogeley}
\affil{Department of Physics, Drexel University, 3141 Chestnut Street, Philadelphia, PA  19104}
\email{kelly.a.douglass@drexel.edu}

\begin{abstract}
We study how the cosmic environment affects galaxy evolution in the Universe by 
comparing the metallicities of dwarf galaxies in voids with dwarf galaxies in 
more dense regions.  Ratios of the fluxes of emission lines, particularly those 
of the forbidden [\ion{O}{3}] and [\ion{S}{2}] transitions, provide estimates of 
a region's electron temperature and number density.  From these two quantities 
and the emission line fluxes [\ion{O}{2}] $\lambda 3727$, [\ion{O}{3}] 
$\lambda 4363$, and [\ion{O}{3}] $\lambda \lambda 4959,5007$, we estimate the 
abundance of oxygen with the Direct $T_e$ method.  We estimate the metallicity 
of 42 blue, star-forming void dwarf galaxies and 89 blue, star-forming dwarf 
galaxies in more dense regions using spectroscopic observations from the Sloan 
Digital Sky Survey Data Release 7, as re-processed in the MPA-JHU value-added 
catalog.  We find very little difference between the two sets of galaxies, 
indicating little influence from the large-scale environment on their chemical 
evolution.  Of particular interest are a number of extremely metal-poor dwarf 
galaxies that are less prevalent in voids than in the denser regions.
\end{abstract}


\maketitle

\section{Introduction}

Galaxy redshift surveys have shown that the large-scale structure of the galaxy 
distribution is similar to that of a three-dimensional cosmic web \citep{Bond96} 
in which the voids (large, underdense regions that fill upwards of 60\% of 
space) separate galaxy clusters connected by thin filaments of galaxies.  The 
voids found in early surveys \citep[e.g.,][]{Gregory78, Kirshner81, 
deLapparent86} proved to be an ubiquitous feature of large-scale structure.  
Analyses of the Sloan Digital Sky Survey \citep{Abazajian09, Ahn12} have yielded 
catalogs of $10^3$ voids \citep{Pan12, Sutter14}.

These cosmic voids are an important environment for studying galaxy formation 
(see \cite{vandeWeygaert11} for a review).  Gravitational clustering within a 
void proceeds as if in a very low density universe, in which aggregation of 
gravitationally bound dark matter halos ends relatively early and there is 
relatively little subsequent interaction between galaxies, both because of the 
lower density and the faster local Hubble expansion.  Thus, the $\Lambda$CDM 
cosmology predicts that galaxies formed in voids should have lower mass and may 
be retarded in their star formation when compared to those in more dense 
environments \citep[e.g.,][]{Gottlober03, Goldberg05, Cen11}.  \cite{Goldberg04} 
show that the interior of a spherical void with 10\% of the mean density in a 
$\Omega_{matter} = 0.3$, $h = 0.7$ universe evolves dynamically like an 
$\Omega_{matter} = 0.02$, $\Omega_{\Lambda} = 0.48$, $h = 0.84$ universe.  
Hydrodynamical cosmological simulations by \cite{Cen11} show that the gas in 
voids remains below the critical entropy threshold, allowing the void galaxies 
to continue forming stars.  While the more dense environment of cluster galaxies 
drastically alters their chemical composition and future evolution through the 
relatively frequent occurrences of mergers, tidal stripping, and/or ram-pressure 
stripping, void galaxies evolve in a relatively pristine environment where 
interactions are far less frequent and star formation may proceed up to the 
present epoch because void galaxies are able to retain their gas.

The effects of the void environment should be most obvious in the dwarf 
galaxies.  Dwarf galaxies are sensitive to many astrophysical effects, including 
cosmological reionization, internal feedback from supernova and photoheating 
from star formation, external effects from tidal interactions and ram pressure 
stripping, small-scale details of dark matter halo assembly, and properties of 
dark matter. Many of these effects have been invoked to attempt to resolve the 
discrepancy between the mass function of galaxy halos predicted by $\Lambda$CDM 
and the observed, much smaller density of dwarf galaxies observed in voids (see, 
e.g., \cite{Kravtsov09} for a review).  It is critical to explore dwarfs in 
voids to complement studies of dwarfs in groups and clusters because the 
assembly histories of low-mass galaxies are predicted to be very different 
\citep[e.g.,][]{Gao07, Lackner12} and observations to date show that the 
properties of dwarfs vary dramatically with environment \citep[e.g.,][]{Ann08, 
Geha12}.  Diffuse cold-mode accretion, rather than mergers, has been suggested 
to be the dominant mechanism for growing dark matter halos in voids 
\citep[e.g.,][]{Keres05, Fakhouri09}.  Late-time gas accretion may be possible 
in voids if void galaxies retain a baryonic reservoir up to the present epoch.  
Thus, these few, lonely, faint galaxies test important features of the structure 
formation model and our understanding of galaxy formation ``gastrophysics.''

Observational studies of void galaxies have included examination of photometric 
properties such as luminosity \citep{Hoyle05, Croton05, Moorman15}, color and 
morphological type \citep{Grogin00, Rojas04, Patiri06, Park07, 
vonBendaBeckmann08, Hoyle12} star formation rates estimated from optical 
spectroscopy and UV photometry \citep{Rojas05, Moorman15, Beygu16}, and gas 
content \citep{Kreckel12, Moorman16, Jones16}.  Void galaxies tend to be of 
lower luminosity, of late morphological type, blue, have relatively high rates 
of star formation per stellar mass, and gas rich.

Another important diagnostic of galaxy formation is metallicity, which is a 
measure of the integrated star formation history and is frequently characterized 
by the ratio of the oxygen to hydrogen atomic density (often 
$Z = 12 + \log (\text{O}/\text{H})$, though sometimes given in units of the 
solar metallicity, $Z/Z_{\astrosun}$).  The metallicity should depend on the 
galaxy's star formation history, specifically the percentage of the galaxy's gas 
that has been processed in stars \citep{Guseva09}.  If void galaxies have only 
recently started forming stars or have recently accreted unprocessed gas, we 
would expect these galaxies to have a lower metallicity than those in more dense 
regions (whose star formation started earlier due to e.g., tidally-triggered 
star formation).  Furthermore, gas-phase metallicity is affected by the 
evolution of a galaxy's stellar population and the composition of its 
interstellar medium (ISM).  It reveals a galaxy's history of releasing metals 
into the ISM via supernovae and stellar winds, ejecting gas via galactic 
outflows, and accreting gas from the surrounding environment 
\citep[see, e.g.,][and references therein]{Cooper08,Cybulski14,Hirschmann14}.  
Understanding the evolution of metallicity in galaxies is therefore crucial in 
uncovering the details of galactic evolution.

Observations by \cite{Cooper08, Deng11, Filho15, Pustilnik06, Pustilnik11a, 
Pustilnik11b, Pustilnik13, Pustilnik14} appear to support the hypothesis of 
lower metallicity in void galaxies, while \cite{Kreckel15} find no effect of the 
void environment on their sample of eight void dwarf galaxies.  Most of the 
conclusions of previous work are based on samples containing only a handful of 
galaxies.  Because large sky surveys like SDSS contain a substantial collection 
of dwarf galaxies, we can now analyze the dwarf galaxy population in the 
relatively nearby universe to test this hypothesis with more statistical 
significance.  In particular, the main galaxy sample of SDSS DR7 covers a large 
enough volume to identify over 1000 voids \citep{Pan12} and provides 
spectroscopy to permit metallicity estimates of void dwarf galaxies.  We make 
use of the reprocessed spectroscopic data from the MPA-JHU 
catalog\footnote{Available at \url{http://www.mpa-garching.mpg.de/SDSS/DR7/}} to 
study the metallicity of the large collection of dwarf galaxies in SDSS DR7.  As 
explained by \cite{Tremonti04}, the spectra in the MPA-JHU catalog are analyzed 
with a more detailed stellar continuum, permitting the weaker emission lines to 
become more apparent.  With the dependence of our analysis on weak emission 
lines (especially [\ion{O}{3}] $\lambda 4363$), this detailed treatment of the 
weak emission lines should produce more accurate results.  We study the 
metallicity of these galaxies as a function of large-scale environment, testing 
the hypothesis that void dwarf galaxies have lower gas-phase metallicities than 
dwarf galaxies in more dense regions.

Our paper is organized as follows.  Section 2 describes the theory and method 
for using various emission lines to estimate the metallicity of galaxies.  We 
review the source of our data and errors in Section 3.   Section 4 includes the 
results of our metallicity calculations, and we discuss the likelihood of any 
large-scale environmental influence on these results in Section 5.  Finally, 
Section 6 summarizes our conclusions and discusses future work.

\section{Estimation of galaxy metallicity from optical spectroscopy}
\label{sec:Theory}

\subsection{Overview of Methods}

We characterize the galaxy metallicity using oxygen because it is relatively 
abundant, it emits strong lines for several ionization states in the optical 
regime, and a ratio of its lines provides a good estimate of the electron 
temperature \citep{Kewley02}.  Here, we describe the theory and method we employ 
to estimate oxygen abundances in dwarf galaxies.

UV photons from young stars in an \ion{H}{2} region keep the interstellar gas 
partially ionized.  Optical photons are either absorbed and re-emitted 
throughout the region at resonant frequencies (resulting in classically 
permitted electron transitions), or the electrons are collisionally excited 
(resulting in classically forbidden electron transitions).  Collisional 
excitation of the lower energy levels of metal ions is possible because these 
levels are only a few eV above the ground state \citep{DeRobertis87}.  
Consequently, the UV-optical spectrum contains some of the most useful 
diagnostic emission lines.  Due to observational constraints of SDSS DR7 (the 
spectrometer's wavelength range and the signal-to-noise of the resulting 
spectra; see Section \ref{sec:Data}), not all these emission lines are easily 
measured.

Three classes of methods have been developed to estimate the gas-phase 
metallicity of a galaxy, which we label as direct, theoretical, and empirical.  
Direct-$T_e$ methods are based on a measurement of the [\ion{O}{3}] 
$\lambda 4363$ auroral line, from which a ``direct'' estimate of the electron 
temperature can be made \citep[e.g.,][]{Izotov06, Kniazev08, Pilyugin07, Yin07}.  
Theoretical methods are based on photoionization models 
\citep[e.g.,][]{Kewley02}.  Empirical methods make an indirect estimate of the 
electron temperature based on calibrated relationships between direct 
metallicity estimates and other strong-line ratios in \ion{H}{2} regions 
\citep[see, for example,][]{Pettini04, Pilyugin11, Dopita13, LaraLopez13, 
Marino13}.  While each of these methods provides an estimate for the 
metallicity, they are all developed for use on sets of galaxies with different 
characteristics (stellar mass or gas-phase metallicity, for example).  
Previously, most theoretical and empirical methods have been calibrated with 
galaxies of larger stellar mass and higher luminosity.  Because the properties 
of dwarf galaxies differ from those of higher luminosity (and larger stellar 
mass), most of these methods drastically over- or under-estimate the gas-phase 
metallicity for dwarf galaxies.  Consequently, we must exercise caution when 
applying these various calculation methods for estimating the metallicity of 
dwarf galaxies.  We attempt to avoid any calibration issues by estimating 
gas-phase metallicity using the direct-$T_e$ method.  This method relies on the 
weak [\ion{O}{3}] $\lambda 4363$ emission line, which limits the number of dwarf 
galaxies we can analyze.  However, because this method provides more reliable 
metallicity estimates than any of the others for dwarf galaxies, we chose 
quality over quantity in our results.

\subsection{[\ion{O}{3}]}\label{sec:O3}

\begin{figure}
	\centering
    \includegraphics[width=0.47\textwidth]{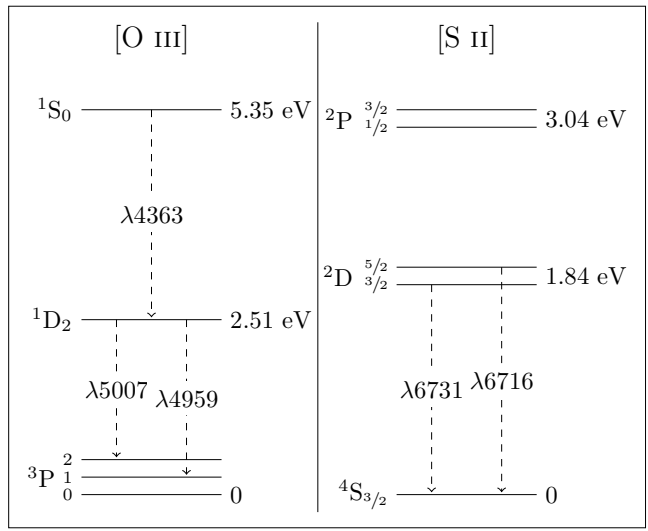}
	\caption{Energy-level diagram for [\ion{O}{3}] $(2p^2)$ and [\ion{S}{2}] 
	$(3p^3)$ ions.  The most important transitions are shown; all are in the 
	visible spectrum.  These forbidden transitions in oxygen provide an estimate 
	of the electron temperature in the interstellar gas, while the forbidden 
	sulfur transitions provide an estimate of the electron number density.  With 
	estimates of the electron temperature and number density, we can convert 
	emission line flux ratios into chemical abundance ratios.}
	\label{fig:transitions}
\end{figure}

There are three significant emission lines for doubly-ionized oxygen.  The 
relative excitation rates to the $^1$S and $^1$D energy levels depend very 
strongly on the electron temperature, $T_e$; therefore, the relative strengths 
of these emitted lines can be used to measure the electron temperature 
\citep{Osterbrock89}.  In the low-density limit $(n_e < 10^5 \text{ cm}^{-3})$, 
most excitations to the $^1$D level result in an emission of a photon with a 
wavelength of either $5007\text{\AA}$ or $4959\text{\AA}$, as shown in Fig. 
\ref{fig:transitions}.  Most excitations up to $^1$S produce a photon of 
wavelength $4363\text{\AA}$, followed by a photon of either of the two previous 
wavelengths (since the electron is now in the $^1$D level).

At higher densities, collisional de-excitation begins to influence these 
emission rates \citep{Osterbrock89}.  Because the $^1$D level has a longer 
lifetime than the $^1$S state, it is collisionally de-excited at lower electron 
densities.  This weakens the $\lambda 4959$ and $\lambda 5007$ emission lines.  
At the same time, the additional collisional excitations of the $^1$D state 
permitted by the higher electron densities strengthen the $\lambda 4363$ 
emission line.

[\ion{O}{3}] $\lambda 4363$ is a temperature-sensitive forbidden transition line 
of doubly-ionized oxygen that is the preferred line to use when measuring the 
metallicity of galaxies.  Since the most effective cooling channel in these 
\ion{H}{2} regions is oxygen line emission, lower metallicity regions have higher 
temperatures  \citep{Saintonge07}.  Collisional excitations up to this energy 
level are more common at higher temperatures, since there are more electrons 
with the kinetic energy required to excite the O$^{++}$ ion to this energy 
level.  As a result, the line strength of [\ion{O}{3}] $\lambda 4363$ correlates 
with the region's temperature and is therefore anticorrelated with the 
metallicity of the galaxy.  [\ion{O}{3}] $\lambda 4363$ is already one to two 
orders of magnitude weaker than the [\ion{O}{3}] $\lambda \lambda 4959, 5007$ 
doublet, so it is very difficult to obtain an accurate ratio with this line.  It 
is for these reasons that other ``empirical'' relations were developed for 
metallicity calculations, eliminating the need for an electron temperature 
estimate from this emission line.

Given an electron temperature and density, the flux ratio of the [\ion{O}{3}] 
$\lambda \lambda 4959, 5007$ doublet to H$\beta$ provides an abundance estimate 
for doubly-ionized oxygen.

\subsection{[\ion{O}{2}]}

A less temperature-sensitive line than [\ion{O}{3}] $\lambda 4363$, the 
[\ion{O}{2}] $\lambda 3727$ forbidden transition doublet of singly-ionized 
oxygen is often used in metallicity calculations.  With an electron temperature 
and density, its flux provides an estimate of the abundance of singly-ionized 
oxygen.  In SDSS spectra, this line can be observed for objects with a redshift 
greater than 0.02.  However, because dwarf galaxies are inherently faint objects 
($M_r > -17$), they are targeted for spectroscopy in SDSS only out to redshift 
$z\sim 0.03$, thus we can only estimate the metallicity of dwarf galaxies in the 
redshift range $0.02 < z < 0.03$.

\subsection{[\ion{S}{2}]}

Just as we are able to measure the electron temperature from [\ion{O}{3}] 
transitions, we can estimate the electron number density from [\ion{S}{2}] 
transitions.  Below a density of about $100 \text{ cm}^{-3}$, the [\ion{S}{2}] 
$\lambda 6716 / \lambda 6731$ ratio has a weak dependence on the density.  All 
our galaxies fall within this low-density regime, so we assume a low-density 
limit of $n_e = 100 \text{ cm}^{-3}$.

\subsection{Direct $T_e$ method}

\begin{figure}
    \includegraphics[width=0.5\textwidth]{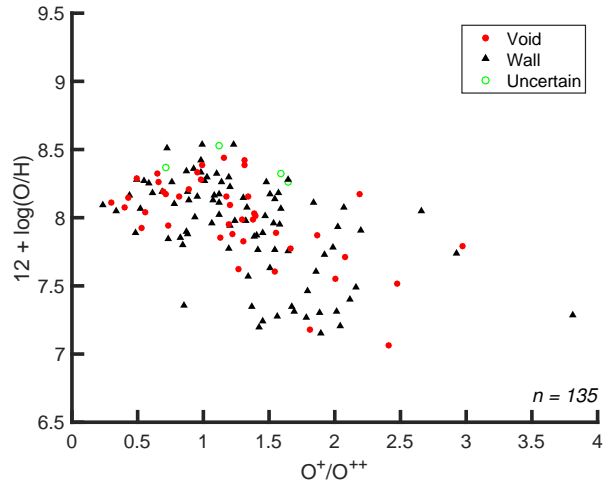}
    \caption{Metallicity of our 135 dwarf galaxies as a function of 
    O$^+$/O$^{++}$.  While either O$^+$ or O$^{++}$ can dominate our galaxies' 
    oxygen abundances, only those with low metallicities (high temperatures) and 
    with O$^+$ dominating the abundance will be affected by the temperature 
    overestimate of the low-ionization zone as found by \cite{Andrews13}.  The 
    small number of galaxies (15/135) that may suffer from this possible 
    temperature overestimate do not affect our results.}
    \label{fig:O2O3_ratio}
\end{figure}

We use the method published by \cite{Izotov06}, which is based on the 
astrophysics in \cite{Osterbrock89}.  It makes use of the [\ion{O}{3}] 
$\lambda 4363$, $\lambda \lambda 4959, 5007$ lines and the [\ion{O}{2}] 
$\lambda 3727$ doublet.  While often regarded as the most accurate estimate of 
the metallicity, it is difficult to employ due to the restrictions on 
[\ion{O}{3}] $\lambda 4363$.  Consequently, this method is best suited for 
low-redshift, low-metallicity galaxies.  The electron temperature is derived by 
solving the following system of equations:
\begin{equation}
	t_3 = \frac{1.432}{\log[(\lambda 4959 + \lambda 5007)/\lambda 4363] - \log C_T}
\end{equation}
where $t_3 = 10^{-4} T_e(\text{O}^{++})$ and
\begin{equation}
	C_T = (8.44 - 1.09t_3 + 0.5t_3^2 - 0.08t_3^3)\frac{1 + 0.0004x_3}{1 + 0.044x_3}
\end{equation}
where $x_3 = 10^{-4} n_e t_3^{-0.5}$.  The ionic abundances are then found with 
the equations
\begin{align}
	12 + \log \left( \frac{\text{O}^+}{\text{H}^+} \right) &= \log \frac{\lambda 3727}{\text{H}\beta} + 5.961 + \frac{1.676}{t_2} \nonumber \\
	&\qquad - 0.40\log t_2 - 0.034t_2 \nonumber \\
	&\qquad + \log (1+1.35x_2) \\
	12 + \log \left( \frac{\text{O}^{++}}{\text{H}^+} \right) &= \log \frac{\lambda 4959 + \lambda 5007}{\text{H}\beta} + 6.200 \nonumber \\
	&\qquad + \frac{1.251}{t_3} - 0.55\log t_3 \nonumber \\
	&\qquad - 0.014t_3
\end{align}
where $t_2 = 10^{-4} T_e(\text{O}^+)$, $t_2 = 0.7t_3 + 0.3$ \citep{Garnett92}, 
and $x_2 = 10^{-4} n_e t_2^{-0.5}$.  \cite{Andrews13} show that this relation 
between $T_e(\text{O}^+)$ and $T_e(\text{O}^{++})$ may overestimate the 
temperature in the low ionization zone, causing the calculated metallicities to 
be underestimated.  Because we care only about the relative metallicity values 
of the galaxies, this effect will only affect our results in galaxies where 
O$^+$ dominates the oxygen abundance (where O$^+$/O$^{++} > 1$) in higher 
temperature regions (or low metallicities).  As shown in Fig. 
\ref{fig:O2O3_ratio}, this affects perhaps fifteen galaxies and does not change 
our conclusions.

The total gas-phase oxygen abundance is equal to the sum of the abundances of 
each of the ionized populations:
\begin{equation}
	\frac{\text{O}}{\text{H}} = \frac{\text{O}^{++}}{\text{H}^+} + \frac{\text{O}^+}{\text{H}^+}
\end{equation}

\section{SDSS data and galaxy selection}\label{sec:Data}

The SDSS Data Release 7 (DR7) \citep{Abazajian09} is a wide-field multi-band 
imaging and spectroscopic survey, using drift scanning to map approximately 
one-quarter of the northern sky.  Photometric data in the five band SDSS system 
--- $u$, $g$, $r$, $i$, and $z$ --- are taken with a dedicated 2.5-meter 
telescope at the Apache Point Observatory in New Mexico \citep{Fukugita96, 
Gunn98}.  Galaxies with a Petrosian $r$-band magnitude $m_r < 17.77$ are 
selected for spectroscopic analysis \citep{Lupton01, Strauss02}.  The spectra 
have an observed wavelength range of $3800\text{\AA}$ to $9200\text{\AA}$ with a 
resolution $\lambda / \Delta \lambda \sim 1800$, and are taken using two 
double fiber-fed spectrographs and fiber plug plates with a minimum fiber 
separation of 55 arcseconds \citep{Blanton03}.  The emission line flux data used 
in this study are from the MPA-JHU value-added catalog, which is based on the 
SDSS DR7 sample of galaxies.  Absolute magnitudes, colors, and all other 
additional data are from the KIAS value-added galaxy catalog \citep{Choi10}.

\subsection{Spectroscopic selection}

To satisfy the needs of our analysis, we make the following cuts to our sample.  
All analyzed galaxies must have relatively recent star formation, since UV 
photons are needed to excite the interstellar gas to produce the required 
emission lines.  As a result, each galaxy must have a star-forming BPT 
classification by \cite{Brinchmann04}.  In addition, because we analyze only 
dwarf galaxies ($M_r > -17$), there is a natural redshift upper limit of 0.03 on 
the samples; dwarf galaxies at higher redshifts are not bright enough to be 
included in the spectroscopic data of SDSS.  For a galaxy to be analyzed, we 
require a minimum $5\sigma$ detection of the H$\beta$ emission line and at least 
a $1\sigma$ detection of the [\ion{O}{3}] $\lambda 4363$ forbidden transition.  
The restriction on both these lines eliminate those galaxies with a low S/N 
spectrum.  This is particularly important for [\ion{O}{3}] $\lambda 4363$, as it 
is inherently a weak emission line.  We are aware that implementing this 
restriction on [\ion{O}{3}] $\lambda 4363$ eliminates those galaxies with higher 
metallicities, since the strength of this line is inversely proportional to the 
metallicity of the galaxy (see Sec. \ref{sec:O3} for details).  However, we show 
that this restriction does not affect our conclusions on the large-scale 
environmental dependence on the gas-phase metallicity.

In addition, we also eliminate galaxies with temperature estimates 
$T_e (\text{\ion{O}{3}}) > 3\times 10^4\text{ K}$.  Gas temperatures above this 
threshold are not physical for an \ion{H}{2} region 
\citep[inferred from][]{Osterbrock89, Izotov06, Luridiana15}.

For the dwarf galaxies in our sample, the [\ion{O}{2}] $\lambda 3727$ spectral 
line is very close to the edge of the spectrometer due to their maximum redshift 
$z < 0.03$.  Consequently, its flux measurement is not always reliable.  
Therefore, the flux values labeled \texttt{oii\_flux} in the MPA-JHU catalog are 
used instead of the combined flux values measured for the [\ion{O}{2}] 
$\lambda \lambda 3726,3729$ doublet.  Because the velocity dispersion is not 
fixed when measuring the flux found in \texttt{oii\_flux}, the resulting 
measurements tend to be more realistic than those measured with the fixed 
dispersion (C. Tremonti, private communication).  In addition, those galaxies 
with remaining erroneous measurements for the [\ion{O}{2}] $\lambda 3727$ 
doublet were removed by hand, after comparing the listed flux values to the 
spectra by eye.  All spectral lines used in the analysis must have a flux 
greater than 0, to ensure that they are emission lines.

\subsection{Void classification}

Void galaxies are identified using the void catalog compiled by \cite{Pan12}, 
which was built based on the galaxies in SDSS DR7 catalog.  Starting with 
galaxies with absolute magnitudes $M_r < -20$, the VoidFinder algorithm of 
\cite{Hoyle02} removes all isolated galaxies (defined as having the third 
nearest neighbor more than 7 $h^{-1}$ Mpc away).  After applying a grid to the 
remaining galaxies, spheres are grown from all empty grid cells (cells 
containing no galaxies).  A sphere reaches its maximum size when it encounters 
four galaxies on its surface.  To be classified as a void (or part of one), a 
sphere must have a minimum 10 Mpc radius.  If two spheres overlap by more than 
10\%, they are considered part of the same void.  See \cite{Hoyle02} for a more 
detailed description of the VoidFinder algorithm.  Those galaxies that fall 
within these void spheres are classified as void galaxies.  Those galaxies that 
lie outside the spheres are classified as wall galaxies.  Because we cannot 
identify any voids within 10 Mpc of the edge of the survey, we do not include 
the galaxies that fall within this region in either the void or wall sample 
(throughout this paper, these galaxies are labeled as ``Uncertain'').

Of the $\sim$ 800,000 galaxies with spectra available in SDSS DR7, 9519 are 
dwarf galaxies.  Applying the spectroscopic cuts, 42 void dwarf galaxies, 89 
wall dwarf galaxies, and 4 dwarf galaxies with uncertain large-scale 
environments are left to analyze (for a total of 135 dwarf galaxies, 131 of 
which are used in the environmental tests).

\section{Metallicity analysis and results}

Our primary objective is to perform a relative measurement of metallicity of 
dwarf galaxies to discern how the large-scale environment affects their chemical 
evolution.  As discussed in Section \ref{sec:Theory}, the strength of and 
ability to observe different spectral lines between various surveys and 
observations require multiple methods to be developed for metallicity 
calculations.  In this paper, we use only the Direct $T_e$ method, because no 
other method has yet been calibrated using dwarf galaxies.  The results from the 
various methods are not directly comparable; while they all return metallicities 
within the same range, the same galaxy can have very different metallicity 
values depending on which method is used.  Conversions between methods have been 
developed \citep[see][]{Kewley08}, but it is not clear that these conversions 
would be accurate for dwarf galaxies.  Unfortunately, there are not enough 
galaxies available in our sample to calibrate these other methods for dwarf 
galaxies.  

All line ratios listed are ratios of the emission line fluxes.  Galaxies with 
low metallicities have $Z = 12 + \log (\text{O}/\text{H}) < 7.6$ 
\citep{Pustilnik06}; galaxies with high metallicities have $Z > 8.2$ 
\citep{Pilyugin06}.  The solar metallicity is $Z_{\astrosun} = 8.69\pm 0.05$ 
\citep{Asplund09}.

\subsection{Estimation of uncertainties and confirmation of our method}

We estimate uncertainties in the computed metallicity using a Monte-Carlo 
method.  Using the measured line fluxes and scaled uncertainty 
estimates\footnote{As described at 
\url{http://wwwmpa.mpa-garching.mpg.de/SDSS/DR7/raw_data.html}} from the 
MPA-JHU catalog, 100,000 different metallicities are calculated for a given 
galaxy.  For each estimate, the flux of a line is drawn from a normal 
distribution, with the expectation value being the original measured flux and 
the standard deviation being the given error in the flux measurement.  
We require all simulated line fluxes to be positive, as negative flux values 
would result in erroneous metallicity values.  The standard deviation in the set 
of these 100,000 calculated values is used as the error in the metallicity 
estimate for the galaxy.  As a result, these uncertainties tend to be larger 
than those quoted in other sources, as they include more information than just 
the quality of the fit used to derive the metallicity.

\begin{figure}
    \centering
    \includegraphics[width=0.5\textwidth]{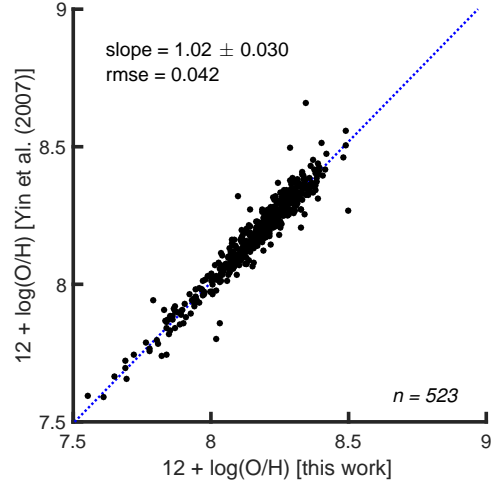}
    \caption{Metallicity ($12 + \log(\text{O}/\text{H})$) comparison between our 
    calculated estimates and those made by \cite{Yin07}.  Error bars have been 
    omitted for clarity.  These are not the dwarf galaxies analyzed in this 
    paper, but rather the sample of galaxies analyzed by \cite{Yin07} to confirm 
    that our version of the calculation is correct.  Both \cite{Yin07} and we 
    have used the metallicity method outlined by \cite{Izotov06}.}
    \label{fig:Yin07_comp}
\end{figure}

We compare results of our analysis of the same set of SDSS galaxies that 
\cite{Yin07} analyze to confirm that our code was working properly, since 
\cite{Yin07} also uses the metallicity method outlined in \cite{Izotov06}.  The 
results of this comparison can be seen in Fig. \ref{fig:Yin07_comp}.  
\cite{Yin07} also uses the MPA-JHU catalog as the source for their data, so our 
results should be identical.

\subsection{Results}

\begin{figure*}
    \centering
    \includegraphics[width=0.49\textwidth]{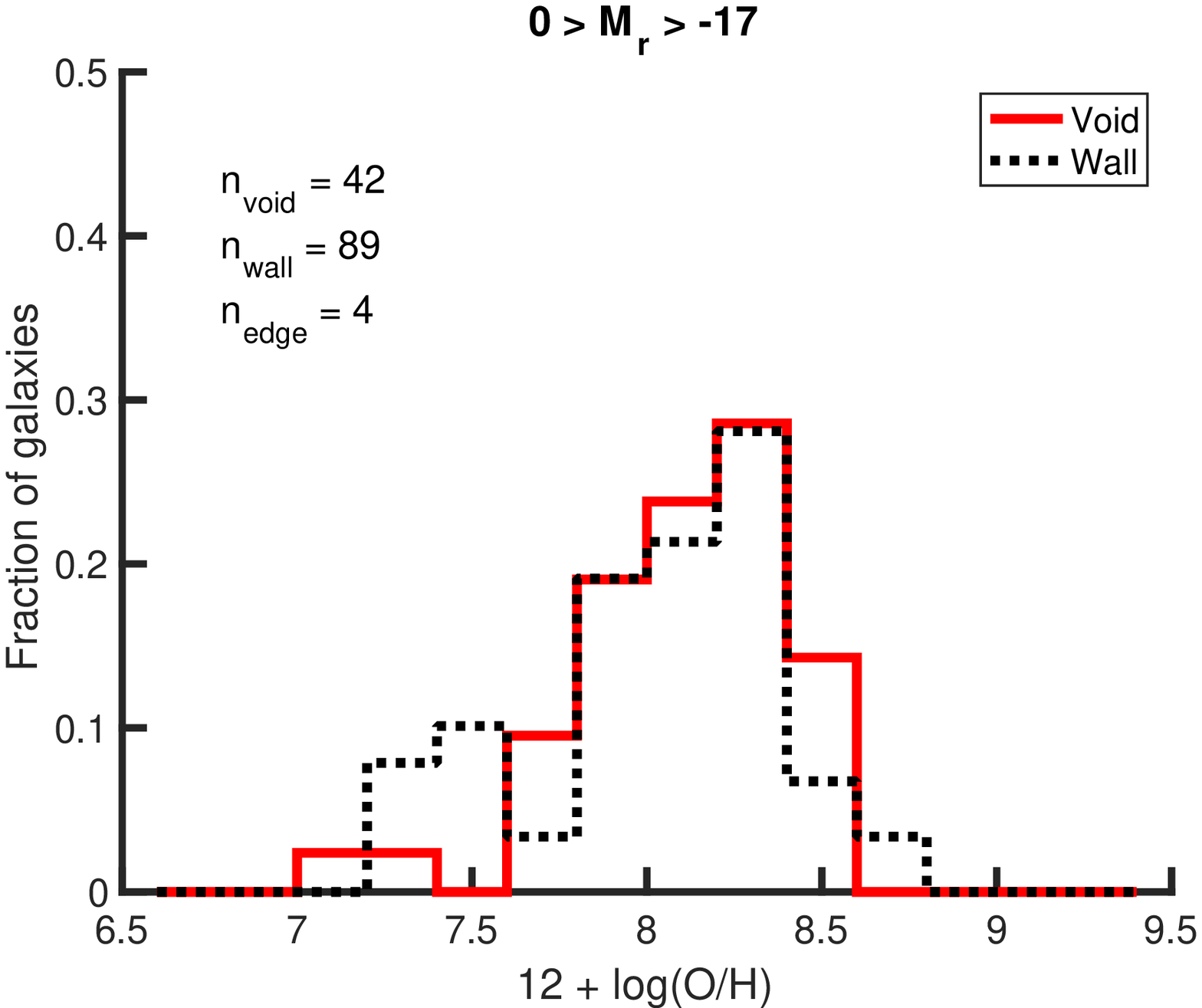}
    \includegraphics[width=0.49\textwidth]{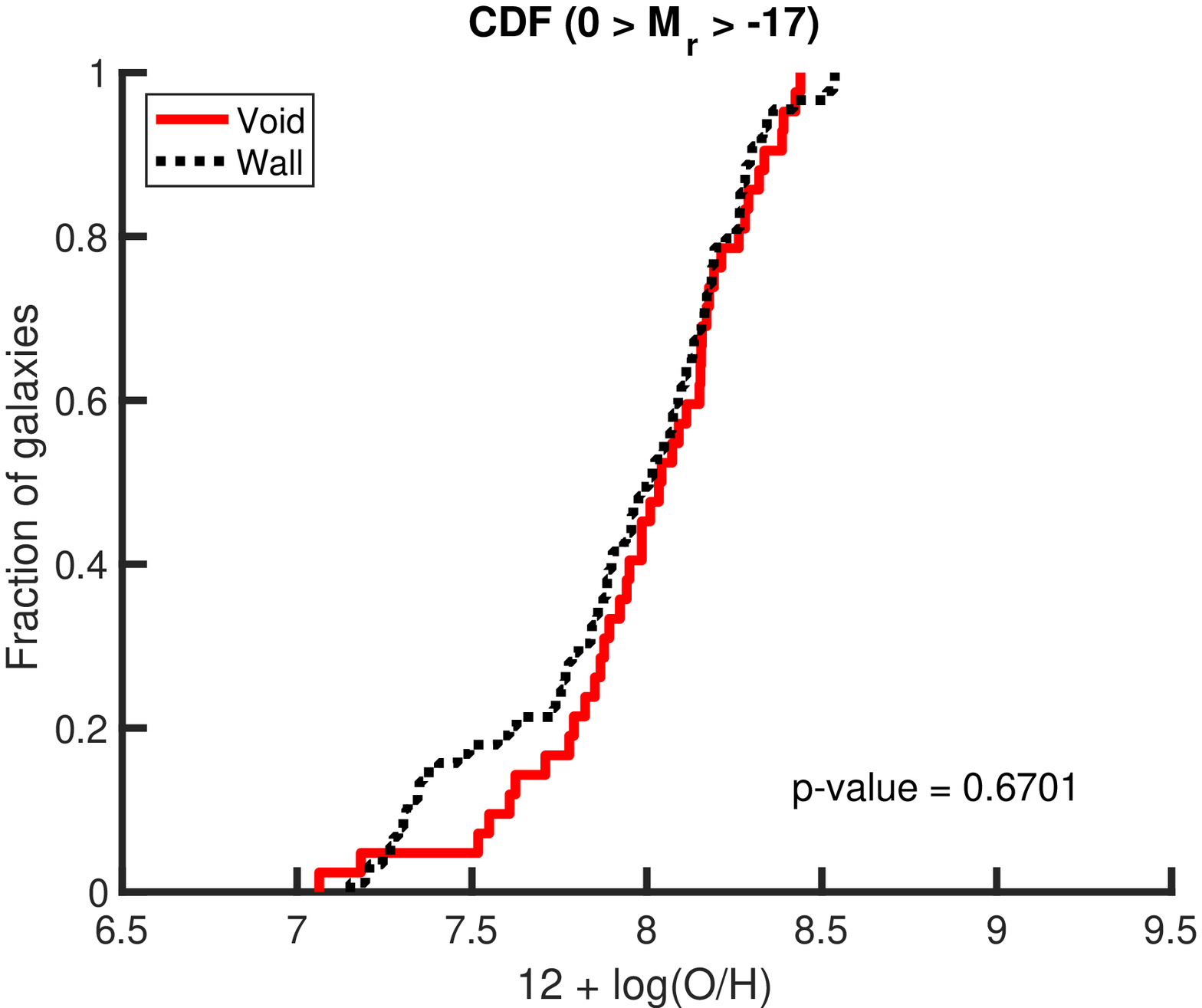}
    \caption{Histogram and associated cumulative distribution function of the 
    gas-phase metallicity of void dwarf (red solid line) and wall dwarf (black 
    dashed line) galaxies.  A two-sample KS test of the two data sets results in an 
    asymptotic $p$-value of 0.67, indicating a 67\% probability that a test 
    statistic greater than the observed value of 0.13 will be seen.  This is 
    reflected visually, as there appears to be very little difference in the two 
    populations, indicating that there is little large-scale environmental 
    influence on the metallicity of dwarf galaxies.}
    \label{fig:met1sig}
\end{figure*}

Metallicities calculated using the Direct $T_e$ method for our dwarf galaxy 
sample are listed in Table \ref{tab:Results}, along with other key 
identification for the galaxies (including whether they are a void or wall 
galaxy).  A histogram of the resulting metallicities is shown in Fig. 
\ref{fig:met1sig}.  As can be seen in Fig. \ref{fig:met1sig}, there is very 
little difference in the spread of metallicity values in dwarf galaxies between 
voids and walls.  A two-sample Kolmogorov-Smimov (KS) test quantifies this 
observation --- it produced a test statistic of 0.13, corresponding to a 
probability of 67\% that a test statistic greater than or equal to that observed 
will be measured if the void sample were drawn from the wall sample; the 
cumulative distribution function (CDF) of these samples can be seen on the right 
in Fig. \ref{fig:met1sig}.

\floattable
\begin{deluxetable}{cccccccc}
\tablewidth{0pt}
\tablecolumns{8}
\tablehead{\colhead{Index\tablenotemark{a}} & \colhead{R.A.} & \colhead{Decl.} & \colhead{Redshift} & \colhead{$M_r$} & \multicolumn{2}{c}{$12 + \log \left( \frac{\text{O}}{\text{H}} \right)$} & \colhead{Void/Wall}}
\tablecaption{Dwarf galaxy properties\label{tab:Results}}
\startdata
63713 & \RA{09}{20}{04}{.27} & -\dec{00}{30}{08}{.97} & 0.0257 & -16.73 & 7.80 & $\pm$0.41 & Wall \\
73537 & \RA{09}{25}{24}{.23} & +\dec{00}{12}{40}{.39} & 0.0250 & -16.94 & 7.94 & $\pm$0.34 & Wall \\
75442 & \RA{13}{13}{24}{.25} & +\dec{00}{15}{02}{.95} & 0.0264 & -16.81 & 7.55 & $\pm$0.35 & Void \\
168874 & \RA{11}{45}{13}{.16} & -\dec{01}{48}{17}{.68} & 0.0273 & -16.99 & 8.16 & $\pm$0.31 & Wall \\
184308 & \RA{09}{39}{09}{.38} & +\dec{00}{59}{04}{.15} & 0.0244 & -16.73 & 7.36 & $\pm$0.43 & Wall\\
\enddata
\tablecomments{Five of the 135 dwarf galaxies analyzed from SDSS DR7.  The flux values for all required emission lines can be found in the MPA-JHU value-added catalog.  Metallicity values are calculated using the direct $T_e$ method, with error estimates via a Monte Carlo method.  The void catalog of \cite{Pan12} is used to classify the galaxies as either Void or Wall.  If a galaxy is located too close to the boundary of the SDSS survey to identify whether or not it is inside a void, it is labeled as Uncertain.  Table \ref{tab:Results} is published in its entirety online in a machine-readable format.  A portion is shown here for guidance regarding its form and content.}
\tablenotetext{a}{KIAS-VAGC galaxy index number}
\end{deluxetable}

\begin{figure*}
    \centering
    \includegraphics[width=0.49\textwidth]{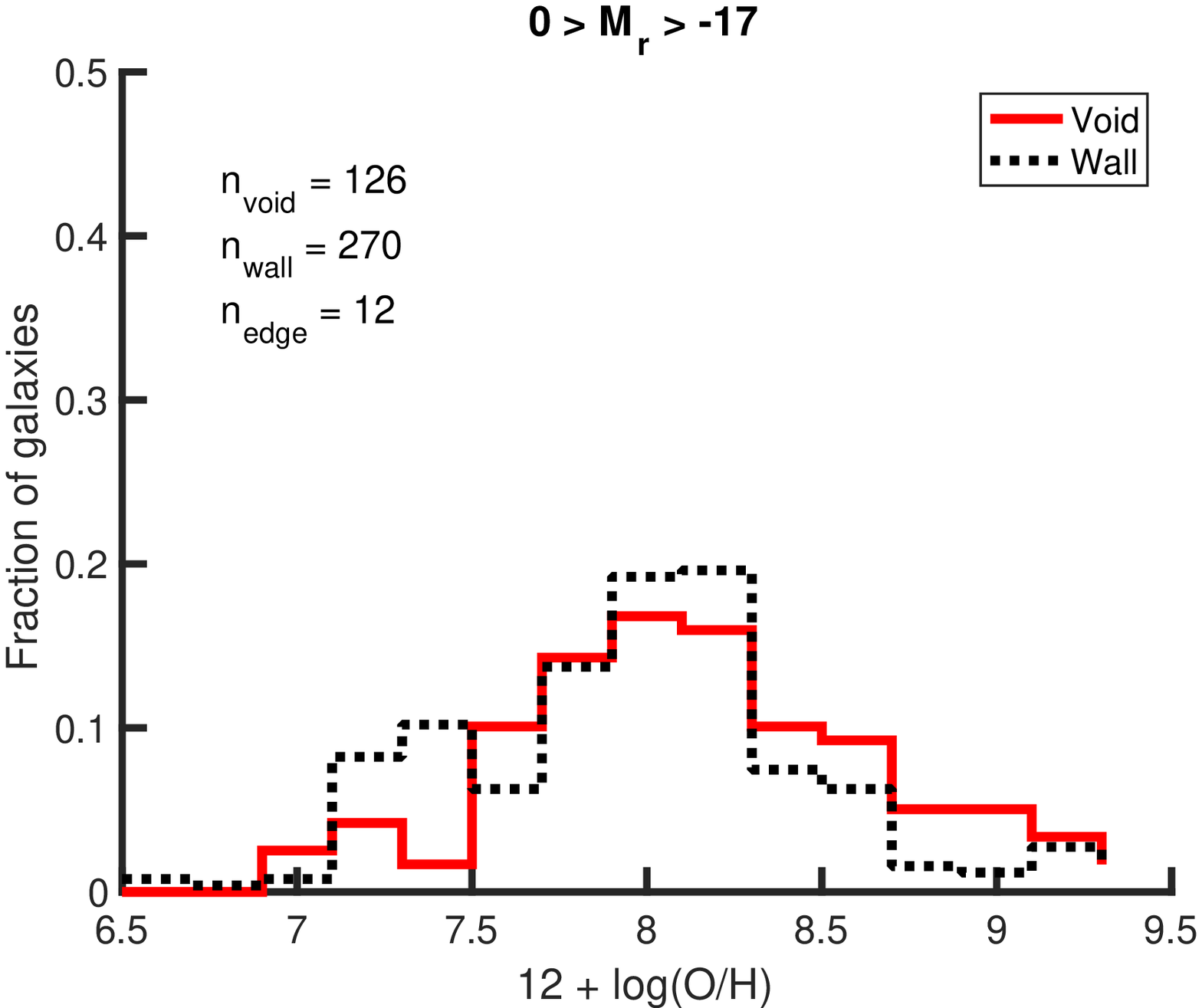}
    \includegraphics[width=0.49\textwidth]{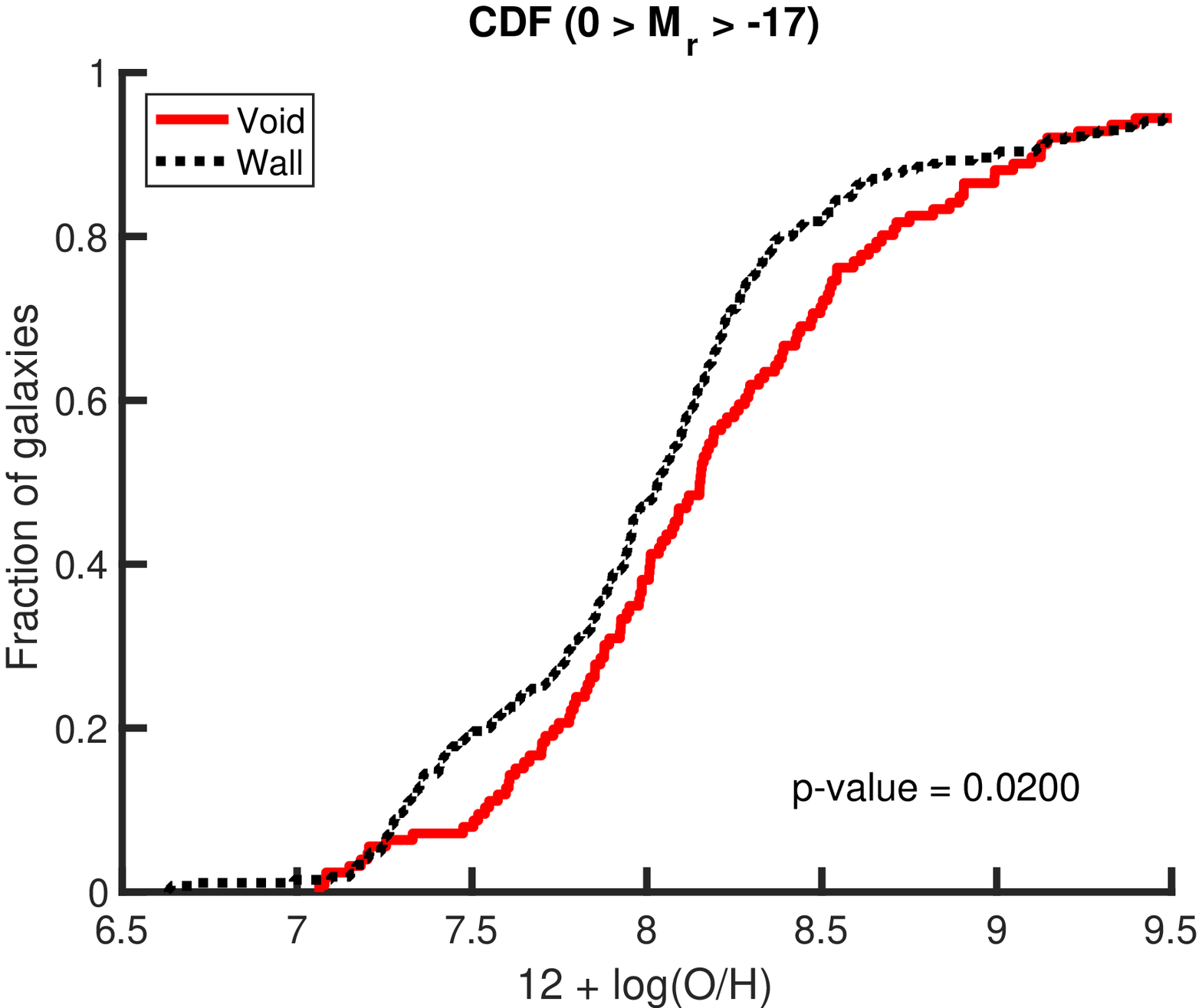}
    \caption{Histogram and associated cumulative distribution function 
    comparing the gas-phase metallicity of void dwarf (red solid line) and wall 
    dwarf (black dashed line) galaxies, testing the effect of the S/N 
    restriction on the auroral [\ion{O}{3}] $\lambda 4363$ line.  The galaxies 
    here have no minimum detection of [\ion{O}{3}] $\lambda 4363$ line.  As 
    expected, eliminating the restriction on this line includes more high 
    metallicity galaxies to the sample, shifting the void dwarf galaxy 
    distribution to have higher metallicities than the wall dwarf galaxies.  
    However, due to the significant uncertainties in the metallicity estimates 
    for $12 + \log (\text{O}/\text{H}) > 8.2$ due to the weak [\ion{O}{3}] 
    $\lambda 4363$ auroral line, this difference in the distributions may not be 
    statistically significant.}
	\label{fig:met0sig}
\end{figure*}

The requirement of a minimum $1\sigma$ detection of [\ion{O}{3}] $\lambda 4363$ 
eliminates galaxies with a low-quality spectrum and those with a weak 
[\ion{O}{3}] $\lambda 4363$ line.  Since this line is inversely proportional to 
the oxygen abundance in the interstellar gas, this biases the sample towards 
more low-metallicity galaxies.  To see how much this cut affects the results, we 
perform the same analysis with no minimum detection limit of [\ion{O}{3}] 
$\lambda 4363$.  As can be seen in Fig. \ref{fig:met0sig}, this adds a 
substantial number of galaxies to the sample (there are now 126 void galaxies 
and 270 wall galaxies analyzed), predominately in the high-metallicity regime.  
As Table \ref{tab:Percents} makes apparent, there is now a higher percentage of 
void dwarf galaxies with high metallicities than wall dwarf galaxies.  However, 
the uncertainties in the metallicity estimates for 
$12 + \log (\text{O}/\text{H}) > 8.2$ are almost 0.5 dex, due to the extremely 
weak [\ion{O}{3}] $\lambda 4363$ auroral line.  Because of these uncertainties, 
the difference in the distributions may not be statistically significant.

\begin{table}
    \centering
    \begin{tabular}{cccc}
        \tableline
         & $Z < 7.6$ & $7.6 \leq Z < 8.2$ & $Z \geq 8.2$\\
        \tableline
        \tableline
        \multicolumn{4}{c}{$1 \sigma$ restriction on [\ion{O}{3}] $\lambda 4363$}\\
        \tableline
        Void & 9.52\% (4)   & 66.67\% (28) & 23.81\% (10)\\
        Wall & 19.10\% (17) & 59.55\% (53) & 21.35\% (19)\\
        \tableline
        \multicolumn{4}{c}{No restriction on [\ion{O}{3}] $\lambda 4363$}\\
        \tableline
        Void & 13.33\% (16) & 45.83\% (55)  & 40.83\% (49)\\
        Wall & 23.26\% (60) & 46.12\% (119) & 30.62\% (79)\\
        \tableline
    \end{tabular}
    \caption{Percentages of galaxies with calculated metallicities within 
    the labeled metallicity ranges, with the number of galaxies in each category 
    in parentheses.  Removing the S/N restriction on [\ion{O}{3}] $\lambda 4363$ 
    especially increases the number of dwarf galaxies with high metallicities, 
    changing the distribution so that void dwarf galaxies have higher 
    metallicities than wall dwarf galaxies.  However, due to the large 
    uncertainties in the metallicity estimates for 
    $12 + \log (\text{O}/\text{H}) > 8.2$, this difference in the distributions 
    may not be statistically significant.\label{tab:Percents}}
\end{table}

\subsection{Sources of systematic error}

It is well-known that many physical properties of galaxies vary with the 
distance from the center of the galaxy \citep{Bell00}.  Therefore, a metallicity 
measurement is dependent on the location of the spectroscopic fiber on the 
galaxy.  If not all the light of the galaxy is contained within the fiber of the 
spectrograph, the estimated metallicity will not necessarily be representative 
of a global metallicity value.  Indeed, it has been shown that different parts 
of a galaxy have different metallicity values \citep{Bell00}.  In SDSS, the 
fiber size is 3 arcseconds -- this corresponds to a physical diameter between 
1.29 kpc and 1.93 kpc at redshifts $0.02 < z < 0.03$.  For many of the dwarf 
galaxies, this covers more than 50\% of the galaxy's luminous surface.  The 
fiber is almost always centered on the brightest spot of the galaxy.  For spiral 
and elliptical galaxies, this is often the center of the galaxy.  Since 
the metallicity of the center of a galaxy is often higher than at its edge, 
these metallicity values may be overestimates of the global metallicity.  Many 
dwarf galaxies are irregular galaxies, where the fiber is instead focused on a 
bright \ion{H}{2} region.

Due to the requirements we place on the emission lines for the galaxies, we are 
inherently limiting our sample to only blue, star-forming galaxies.  This is not 
a representative sample of the dwarf galaxy population.  Rather, in this study 
we are only able to comment on the large-scale environmental influence on blue, 
star-forming dwarf galaxies in a narrow redshift range.  Unfortunately, we 
cannot measure the metallicity of red dwarf galaxies with the Direct $T_e$ 
method, since we need the UV photons from young stars to excite the interstellar 
gas.

\subsection{Comparison to previously published metallicity measurements}

\begin{figure}
    \centering
    \includegraphics[width=0.5\textwidth]{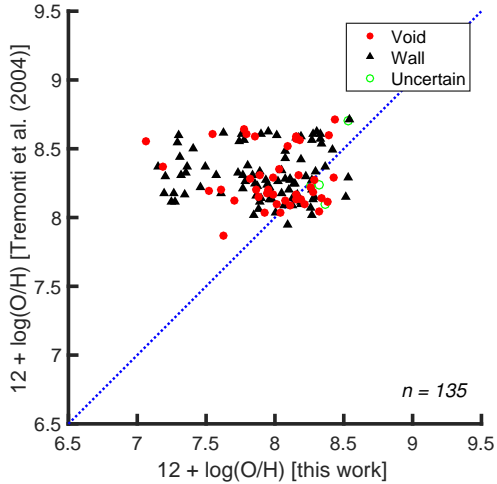}
    \caption{Metallicity ($12 + \log(\text{O}/\text{H})$) comparison between our 
    calculated estimates and those made by \cite{Tremonti04}.  Error bars have 
    been omitted for clarity.  Excepting the extreme low-metallicity galaxies we 
    found, most galaxies agree reasonably well with the values already 
    published.  It is important to note that the strong-line methods 
    \citep[like those used by][]{Tremonti04} are not calibrated for 
    low-metallicity values and are known to overestimate the metallicity by as 
    much as 0.3 dex \citep{Kennicutt03}.  Thus, it is not surprising that oxygen 
    abundances measured using the direct method find lower metallicity, 
    particularly at very low metallicities.}
    \label{fig:T04comp}
\end{figure}

To place our metallicity measurements in the context of previous work, we 
compare our results to the metallicity values measured by \cite{Tremonti04}.  
While we both use data from the MPA-JHU value-added catalog, \cite{Tremonti04} 
employs an empirical method for estimating the metallicity, which is based on 
calibrated relationships between direct metallicity values and strong-line 
ratios.  The results of this comparison are shown in Figure \ref{fig:T04comp}.  
Unfortunately, the range of metallicity values found by \cite{Tremonti04} is 
limited to those galaxies with high metallicities 
$(12 + \log(\text{O}/\text{H}) > 8.5)$, due to the characteristics of their 
sample and their method; they found less than 2\% of their total sample to have 
metallicities less than 8.5.  \cite{Kennicutt03} shows that methods which make 
extensive use of the strong emission lines (so-called ``strong-line'' methods) 
can overestimate the metallicity abundances by as much as 0.3 dex.  A similar 
comparison is made in \cite{Yin07}, where they too find that the metallicity 
estimates of \cite{Tremonti04} are overestimated by 0.34 dex on average.  This 
can be seen quite clearly in Figure \ref{fig:T04comp}, as there is no 
correlation between galaxies with our estimates of 
$12 + \log (\text{O}/\text{H}) < 8$ and the metallicities measured by 
\cite{Tremonti04}, since their metallicities are much higher than ours.  The 
formal correlation coefficient between these two data sets is $0.00 \pm 0.087$; 
the correlation coefficient for those galaxies we measure to have metallicities 
greater than 7.6 (so excluding the low-metallicity galaxies) is 
$0.12 \pm 0.093$.  While this shows a slightly stronger correlation, we realize 
that these galaxies cover a limited range of metallicity values.  As a result, 
any scatter due to the errors in the calculations will result in a low 
correlation coefficient, which is what we see.  Therefore, by Fig. 
\ref{fig:T04comp}, we can see that there is a reasonable agreement between our 
metallicity values and those of \cite{Tremonti04}, excluding those galaxies we 
found to have extremely low metallicity values.

While it is known that there are systematic offsets between different 
metallicity calculation methods \citep{Kewley08}, that does not seem to be the 
case in the relation between our metallicities (measured with the Direct $T_e$ 
method) and those of \cite{Tremonti04} (measured with a combination of 
``strong-line'' methods).  While the metallicity estimates by \cite{Tremonti04} 
do not appear to be significantly biased at 
$8 < 12 + \log (\text{O}/\text{H}) < 8.5$, they overestimate the metallicities 
for low-metallicity galaxies.

\subsection{Mass-metallicity relation}

\begin{figure}
    \centering
    \includegraphics[width=0.5\textwidth]{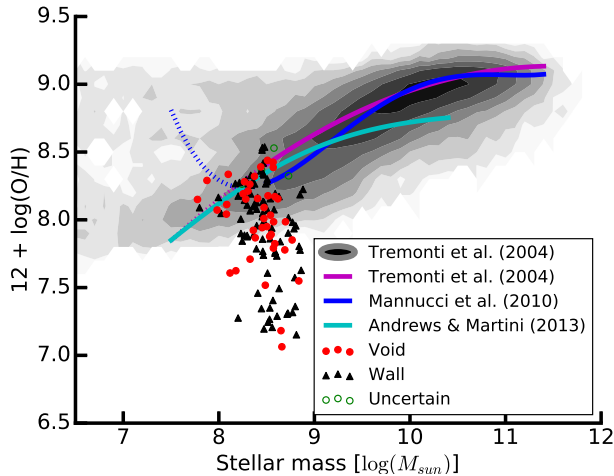}
    \caption{Stellar mass versus metallicity of the 135 analyzed dwarf galaxies.  
    Error bars have been omitted for clarity.  Due to the limited range of mass 
    (all our galaxies are within a small range of masses, since we are looking 
    only at dwarf galaxies), we cannot derive our own relation between the mass 
    and metallicity.  Some previously published relations are plotted over our 
    data for comparison.  To place our sample in context, we have also included 
    (grey contours) those galaxies from the MPA-JHU catalog with metallicity 
    estimates by \cite{Tremonti04}.  It was from these galaxies that the 
    published relation of \cite{Tremonti04} was derived.}
    \label{fig:MZrelation}
\end{figure}

A strong correlation between the stellar mass and metallicity of galaxies 
reflects the fundamental connection between galactic mass and the chemical 
evolution of galaxies.  We use stellar mass estimates from the MPA-JHU catalog 
to examine the mass-metallicity relation in our sample of 135 dwarf galaxies.  
We have also included those galaxies from the MPA-JHU catalog with metallicity 
estimates from \cite{Tremonti04} to place our sample in context.  Due to the 
narrow range of masses in our sample, it is difficult to derive an accurate fit 
to the data.  However, we make comparisons to three published mass-metallicity 
relations \citep{Tremonti04, Mannucci10, Andrews13}.  As can be seen in Fig. 
\ref{fig:MZrelation}, the fit by \cite{Mannucci10} diverges at the low-mass 
limit, and the relations of \cite{Tremonti04} and \cite{Andrews13} predict 
metallicities that are higher than measured for most galaxies in this sample.  
It is important to note that two of these relations are only calibrated down to 
a stellar mass of $10^{8.5} M_{\astrosun}$.  In Fig. \ref{fig:MZrelation}, these 
relations have been extended to $10^{7.5} M_{\astrosun}$, in order to continue 
past our galaxy sample.

In addition to looking at the overall mass-metallicity relation for dwarf 
galaxies, we can also investigate the difference in the relation between 
galaxies in voids and those in more dense regions.  There appear to be no 
significant differences in the two populations, indicating minimal influence 
from the large-scale environment on the mass-metallicity relation of these dwarf 
galaxies.  \cite{Hughes13} also find that the stellar mass-metallicity relation 
is independent of large-scale environment.  This prompts the conclusion that the 
internal evolutionary processes of a galaxy have a greater influence on its 
chemical evolution than its large-scale environment.  We expect this dependence 
of the chemical content of a galaxy on its stellar mass, since the accumulated 
metals reflect the integrated history of star formation.  However, we would 
expect an environmental dependence to appear as well, if void galaxies are in an 
earlier stage of evolution and/or are continuing to accrete fresh gas.

\subsection{SFR-metallicity relation}

\begin{figure*}
    \centering
    \includegraphics[width=0.49\textwidth]{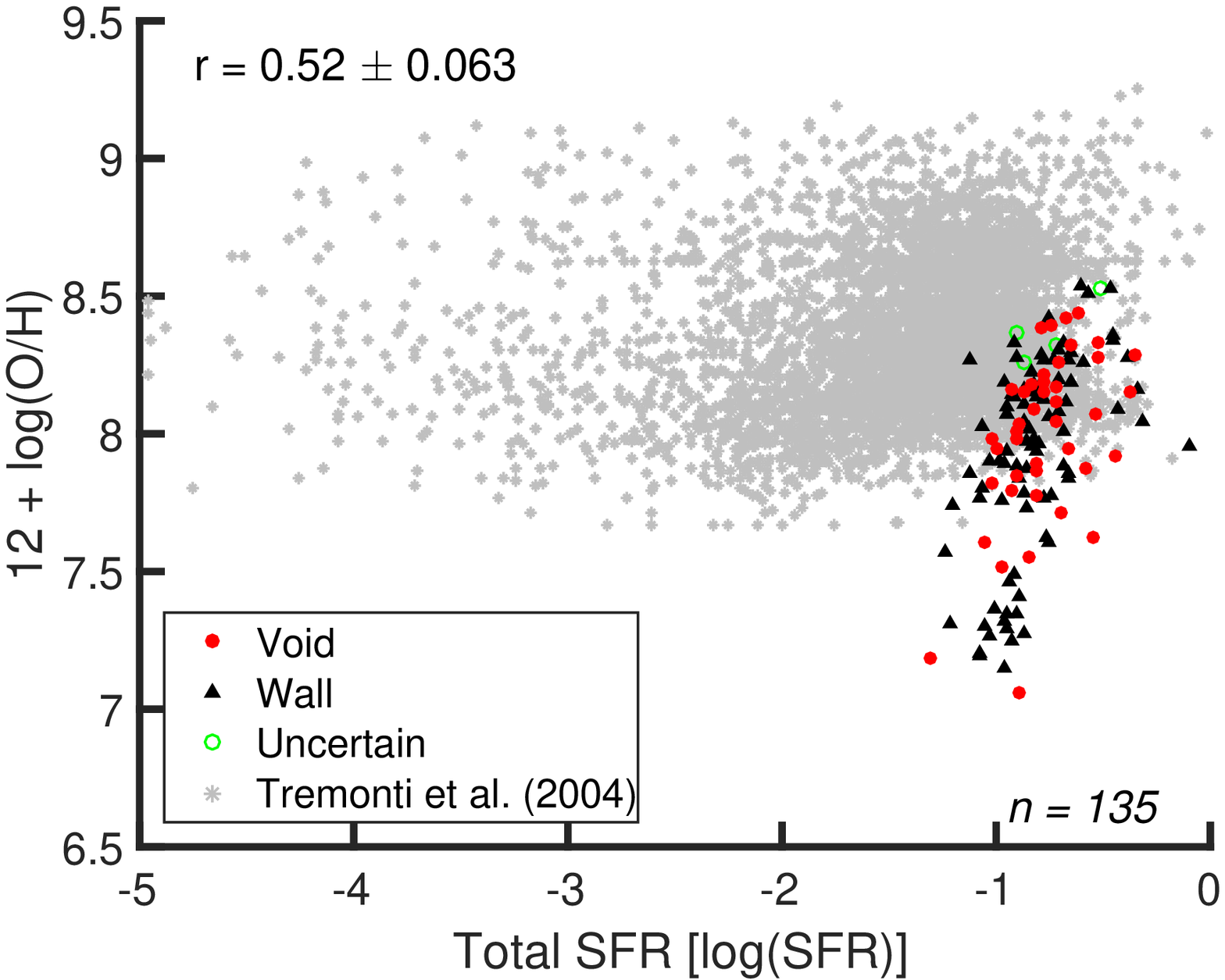}
    \includegraphics[width=0.49\textwidth]{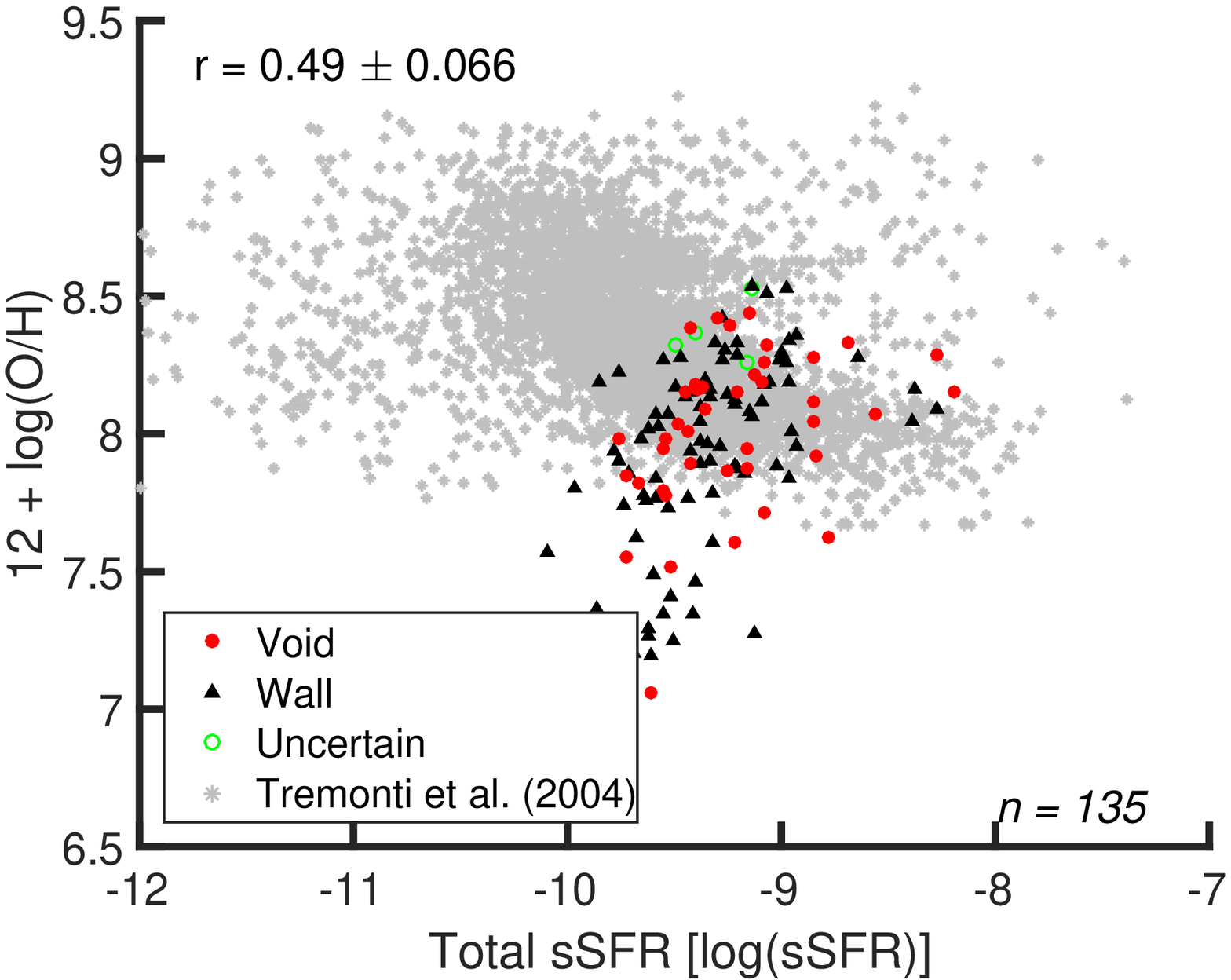}
    \caption{Total star formation rate (SFR) and specific star formation rate 
    (sSFR) versus metallicity of the 135 analyzed dwarf galaxies.  Error bars 
    have been omitted for clarity.  We also plot (grey stars) dwarf galaxies 
    ($M_r > -17$) with metallicity estimates by \cite{Tremonti04}, to place our 
    results in context.  It is significant to note that the majority of our 
    galaxies are on the upper end of the SFR and sSFR for dwarf galaxies, 
    as shown in Fig. \ref{fig:SFR_distribution}.  Note that those galaxies with 
    metallicities $12 + \log(\text{O}/\text{H}) < 7.6$ are on the lower end of 
    the range of sSFR of the dwarf galaxies in our sample.}
    \label{fig:SFRZ_relation}
\end{figure*}

\begin{figure*}
    \centering
    \includegraphics[width=0.49\textwidth]{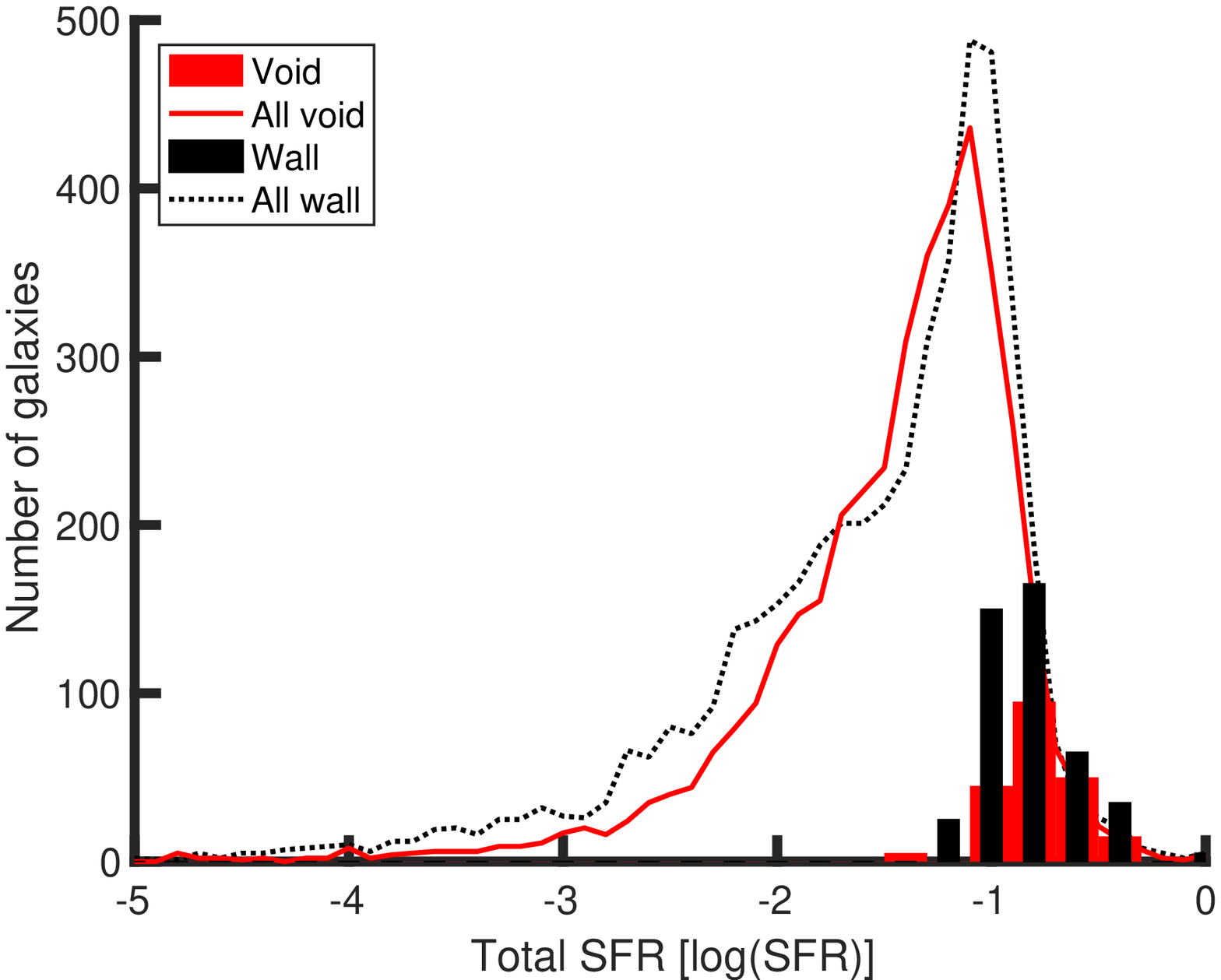}
    \includegraphics[width=0.49\textwidth]{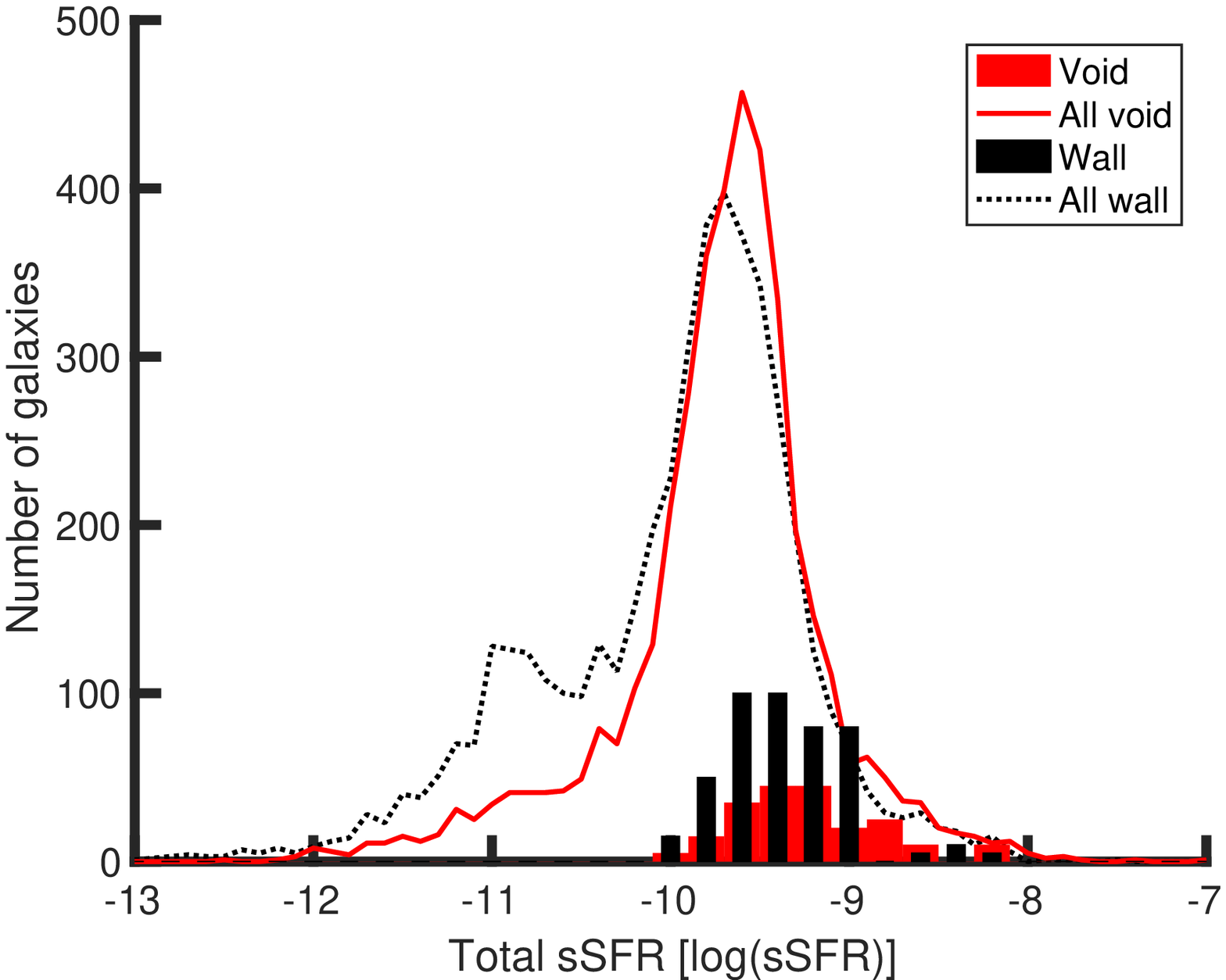}
    \caption{Distribution of the total star formation rate (SFR) and specific 
    star formation rate (sSFR) for void and wall dwarf galaxies in SDSS are 
    shown in the red solid and black dashed lines, respectively.  Our sample of 
    dwarf galaxies (with metallicity values) is shown in the red and black bars 
    (scaled by a factor of 5 for greater visibility).  We are looking only at 
    the highest SFR found in dwarf galaxies; the sSFR for our sample of dwarf 
    galaxies follows the distribution of all dwarf galaxies.  There is clearly a 
    selection bias against lower SFR.}
    \label{fig:SFR_distribution}
\end{figure*}

A fundamental diagnostic of the star formation history of galaxies is the 
relation between stellar mass, metallicity, and star formation rate.  Therefore, 
we also look at the relationship between the (specific) star formation rate and 
metallicity of these 135 dwarf galaxies.  The total (specific) star formation 
rate estimates for these galaxies are from the MPA-JHU value-added catalog, 
based on the technique discussed in \cite{Brinchmann04}.  For low-mass galaxies, 
\cite{Henry13} show that the metallicity is inversely proportional to the star 
formation rate of a galaxy.  However, this is not what is observed in our data, 
as seen in Fig. \ref{fig:SFRZ_relation}.  The correlation coefficient between 
the total (specific) star formation rate and the metallicity 
$r_{sSFR} = 0.49 \pm 0.066$ and $r_{SFR} = 0.52 \pm 0.063$, showing a positive 
correlation between the two properties.  Indeed, those galaxies with the lowest 
metallicities have some of the lowest (specific) star formation rates among the 
dwarf galaxies in our sample.  Since we are limiting our sample to only 
star-forming galaxies, the (s)SFR must be relatively high to emit the UV photons 
needed to ionize the gas.  As a result, all low (s)SFR galaxies will be 
eliminated from our sample, as seen in Fig. \ref{fig:SFR_distribution}.  In 
addition, due to the behavior of the [\ion{O}{3}] $\lambda 4363$ auroral line, 
all galaxies with metallicities $12 + \log (\text{O}/\text{H}) \gtrsim 8.5$ are 
also eliminated from the sample.  As a result, we are only calculating the 
metallicity of galaxies in the lower right corners of the (s)SFR plots in Fig. 
\ref{fig:SFRZ_relation}, which is why we see the unexpected correlation.  There 
does not seem to be a difference between the void and wall galaxies in this 
relation, indicating no large-scale environmental influence on the (s)SFR--Z 
relation.

\subsection{Color-metallicity relation}

\begin{figure*}
    \centering
    \includegraphics[width=0.49\textwidth]{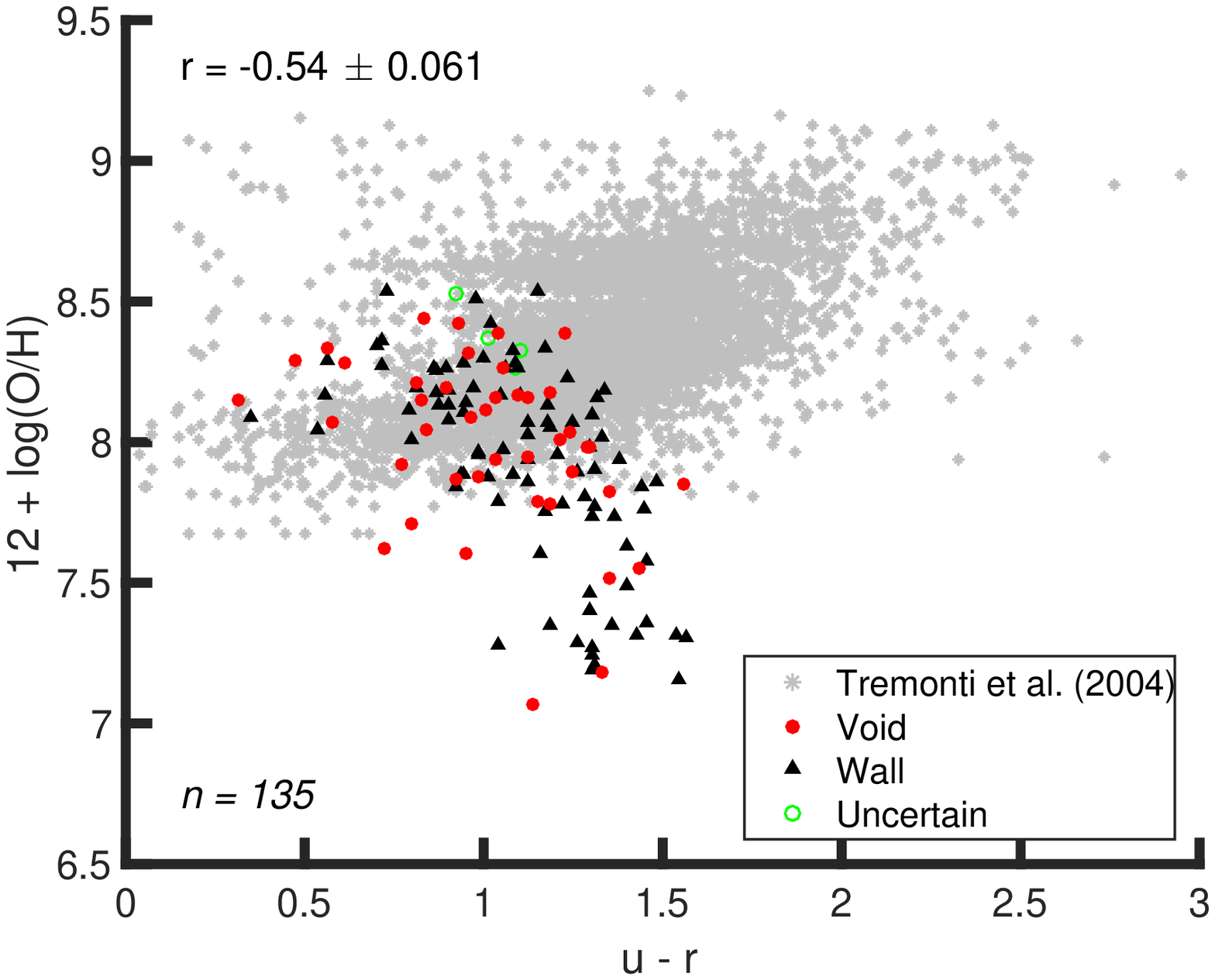}
    \includegraphics[width=0.49\textwidth]{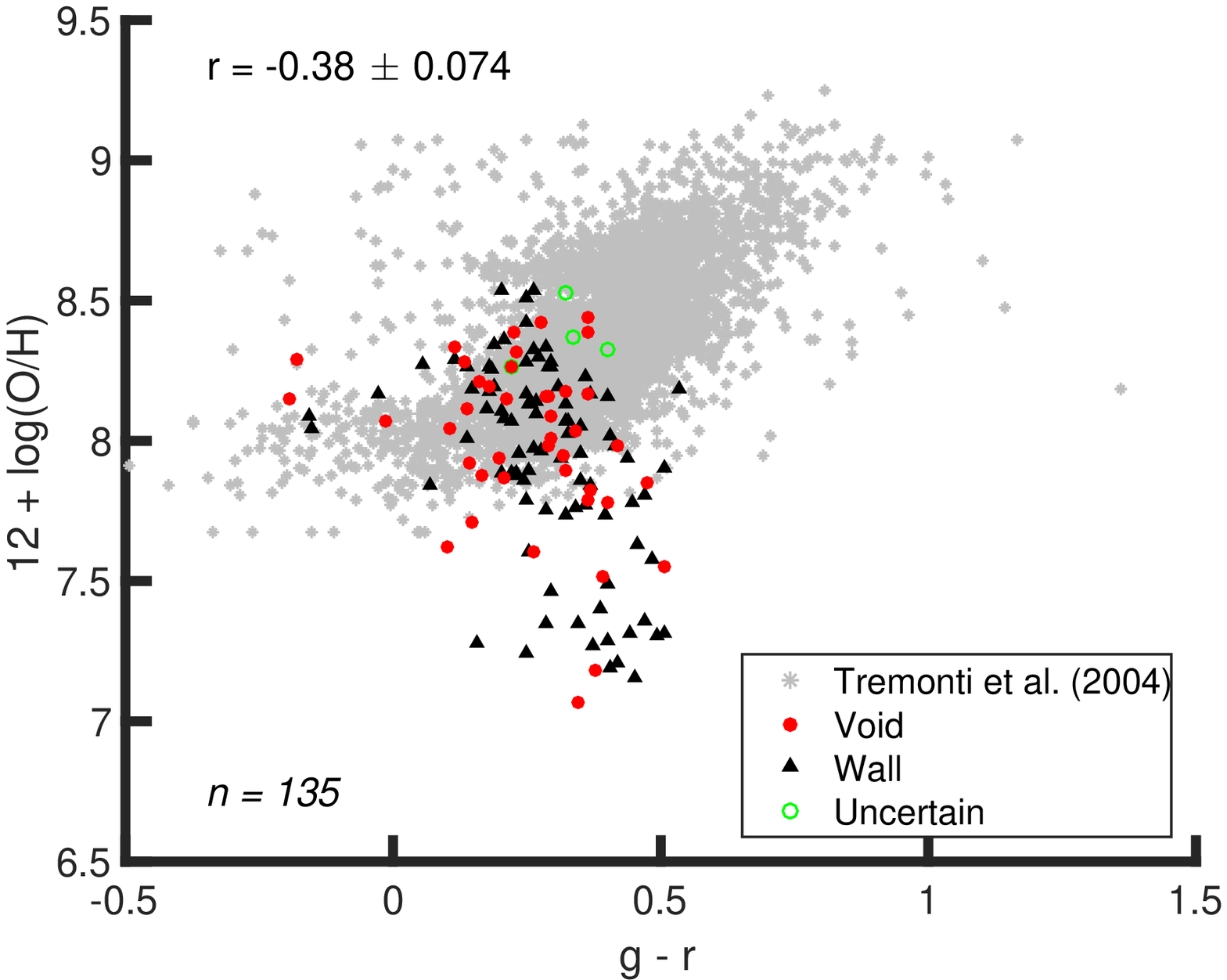}
    \caption{Color (u--r and g--r) versus metallicity of the 135 analyzed dwarf 
    galaxies.  Error bars have been omitted for clarity.  Metallicity is 
    expected to have a positive correlation with color, as older galaxies are 
    expected to have higher metallicities.  To place our galaxies in the context 
    of the dwarf galaxy population, we also plot (grey stars) dwarf galaxies 
    ($M_r > -17$) with metallicity estimates by \cite{Tremonti04}.  We find no 
    significant difference between the void and wall dwarf galaxies, indicating 
    little to no large-scale environmental influence on the color-metallicity 
    relation.}
    \label{fig:colorZ_relation}
\end{figure*}

Metallicity is expected to have a positive correlation with color, as older 
galaxies are expected to have higher metallicities, since they have had more 
time to convert their gas into heavier elements through star formation.  
Therefore, we also look at the color--metallicity relation of our sample of 
135 galaxies -- these relations can be seen in Fig. \ref{fig:colorZ_relation}.  
To place our galaxies in the context of other dwarf galaxies, we have included 
the sample of dwarf galaxies for which \cite{Tremonti04} has estimated 
metallicities (grey stars in the figures).

\begin{figure}
    \centering
    \includegraphics[width=0.5\textwidth]{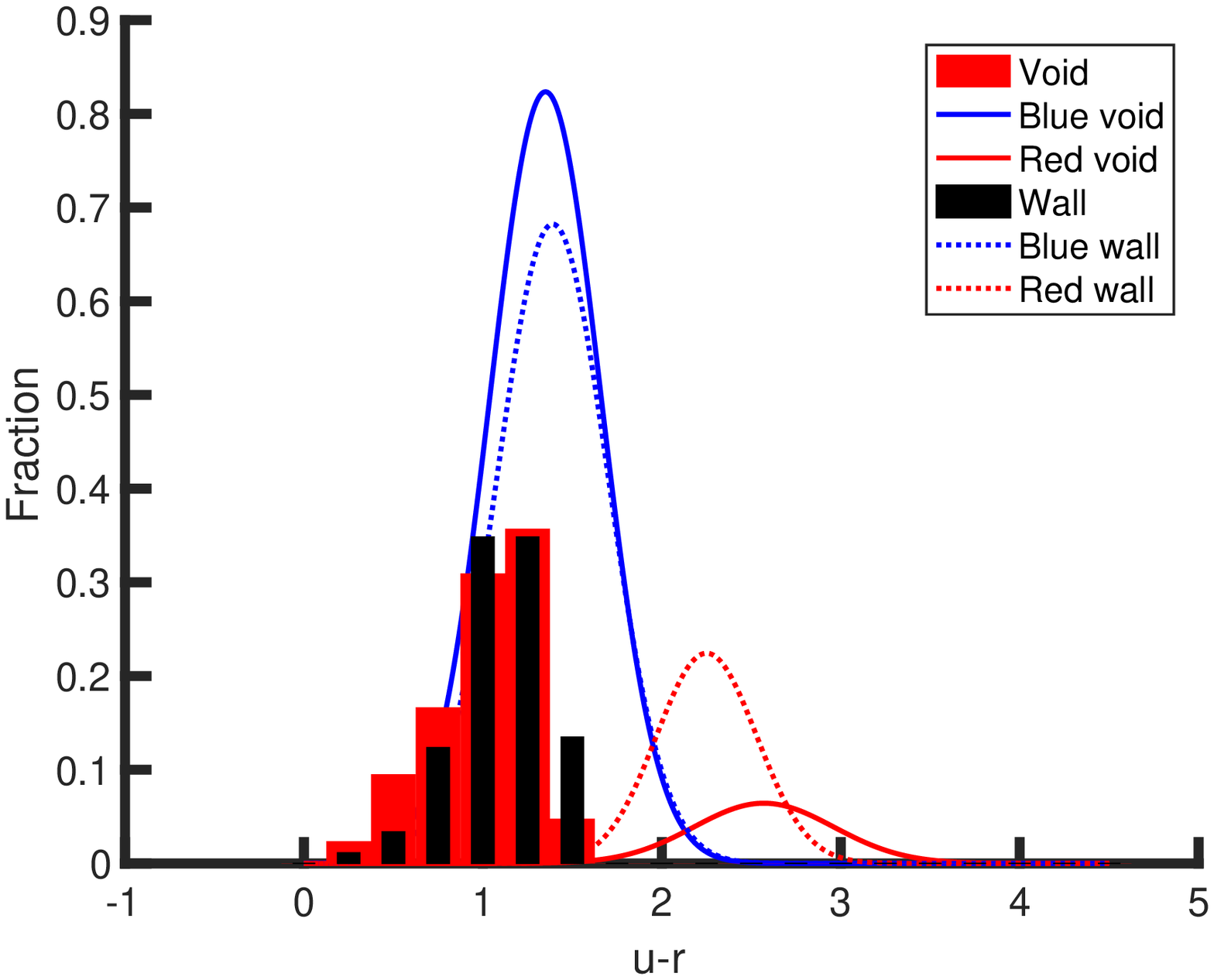}
    \caption{The $u-r$ color distribution of our 135 dwarf galaxies (red/black 
    histograms) as compared to the color distribution of all SDSS dwarf galaxies 
    as found in Fig. 4 of \cite{Hoyle12} (red/blue curves).  It is clear that 
    our galaxies are among the bluest dwarf galaxies in SDSS.}
    \label{fig:colorHist}
\end{figure}

As we can see in Fig. \ref{fig:colorHist}, by overlaying our distribution of 
dwarf galaxies on Fig. 4 of \cite{Hoyle12}, all of our dwarf galaxies are 
members of the blue dwarf galaxy population.  (The Gaussian parameters for the 
curves are taken from Table 3 in \cite{Hoyle12}.)  This is as expected, since 
the Direct $T_e$ method requires measurements of the emission lines of the 
galaxies; these emission lines are caused by the UV photons of newly formed 
stars, indicating a star-forming galaxy and giving the galaxy a blue color.

While the majority of our galaxies follow the positive correlation between color 
and metallicity, the group of extremely low-metallicity galaxies is less blue 
than their metallicities would indicate.  However, when compared to the red/blue 
curves in Fig. \ref{fig:colorHist}, these galaxies occupy the typical range of 
blue dwarf galaxies, so their colors are not unique.  There is no clear 
separation between the void and wall dwarf galaxies in Fig. 
\ref{fig:colorZ_relation}, indicating that there is little or no large-scale 
environmental influence on the color-metallicity relation of these galaxies.

\section{Discussion}

\subsection{Comparison to literature results}

We find no clear distinction between the metallicities of dwarf galaxies in 
voids and dwarf galaxies in more dense regions.  This result agrees with the 
results of \citet{Mouhcine07, Cooper08, Nicholls14, Kreckel15} but disproves our 
initial hypothesis and contradicts the published results of \cite{Pustilnik06, 
Pustilnik11a, Pustilnik14, SanchezAlmeida16}.  \cite{Cooper08} concludes that 
metal-rich galaxies preferentially reside in high-density regions.  Due to our 
requirement on the [\ion{O}{3}] $\lambda 4363$ auroral line, we have very few 
dwarf galaxies with high metallicities.  As a result, we are not able to confirm 
their conclusions. \cite{Deng11} also reports a relationship between environment 
and metallicity.  However, he highlights a large difference in metallicity as a 
function of redshift which correlates with his two samples.  It is possible that 
the dependence he found is actually the result of a systematic dependence on 
redshift in their metallicity calculation.

Many studies suggest that the metallicity of void galaxies should, on average, 
be lower than that of galaxies in more dense regions.  \cite{Mouhcine07} and 
\cite{Cooper08} both perform statistical studies of this relationship on SDSS 
DR4, and \cite{Deng11} repeats this with the DR7 data (only looking at galaxies 
with a redshift $z > 0.02$).  \cite{Mouhcine07} conclude that the relation 
between stellar mass and metallicity is much stronger than that between a 
galaxy's environment and its metallicity.  \cite{Cooper08} find a more 
substantial correlation between a galaxy's environment and its metallicity, but 
point out that the noise of the different methods used to calculate metallicity 
is larger than any environment-metallicity relation.  Our analysis shows that 
there is very little difference between void and wall dwarf galaxies, suggesting 
that the large-scale environment does not strongly influence a dwarf galaxy's 
chemical evolution.

\subsection{Large-scale environmental influence}

Consideration of interactions between the interstellar medium (ISM), 
circumgalactic medium (CGM), and intergalactic medium (IGM) suggests that void 
galaxies should have relatively lower metallicity than galaxies in denser 
environments.  We find no such trend, perhaps because the IGM around 
star-forming ``wall galaxies'' in our sample is similar to that of void 
galaxies.

Simulations by \cite{Cen11} show that the entropy of gas in the IGM in voids 
remains below the critical entropy (defined to be when the cooling time of the 
gas is equal to the Hubble time), so the gas from the IGM can cool and fall into 
a void galaxy's CGM.  In a galaxy's ISM, supernovae expel gas (primarily 
metal-rich) into the CGM.  This gas has a higher metallicity than the average 
metallicity of the ISM \citep[shown by][]{Muratov16}.  While some of this gas 
reaches the outer edge of the CGM, most of it cools and falls back onto the 
galaxy's ISM, after having mixed with the hydrogen that has entered the CGM from 
the IGM.  Therefore, the gas falling back into the galaxy's ISM has a lower 
metallicity than the galaxy's ISM.

In contrast to the void galaxies, the IGM around most wall galaxies is not cool 
enough to fall back onto the CGM.  \cite{Cen11} shows that, in general, the IGM 
of a wall galaxy has an entropy higher than the threshold for cooling.  As a 
result, most of the gas that falls back onto a wall galaxy's ISM is not as 
diluted as what falls onto a void galaxy's ISM.  This is where our hypothesis 
originated: because wall galaxies no longer have a source of cool hydrogen in 
the IGM, their metallicities will be higher than that of the void galaxies (for 
a fixed stellar mass).

However, Fig. \ref{fig:met1sig} does not reveal a lower metallicity in void 
galaxies.  Instead, our results indicate that there is no difference in the 
distribution of metallicities in wall and void galaxies.  In detail, Fig. 10 of 
\cite{Cen11} shows that not all wall galaxies rise above the entropy threshold.  
This is also coincident with the sSFR of the galaxies --- those galaxies with 
higher sSFRs are below the entropy threshold, while those with low sSFRs are 
above (independent of their large-scale environment).  Since all our galaxies 
have relatively high sSFRs (as required by the analysis --- star formation is 
required to detect the emission lines necessary for the metallicity 
calculations), it is possible that our population of wall dwarf galaxies is 
still surrounded by a cool IGM, similar to that of the void galaxies.  As a 
result, the wall galaxies still have a source of cool hydrogen, so the resulting 
distribution of the metallicities in the wall and void dwarf galaxies is the 
same.  \cite{Brisbin12} show that most star-forming galaxies with 
$M_* < 2.0\times 10^{10} M_{\astrosun}$ appear to be fed by the infall of 
pristine or low-metallicity gas.  \cite{Moran12} also find that the lowest-mass 
galaxies ($\log(M_*) < 10.2 M_{\astrosun}$) have a sharp decline in their 
metallicity at large radii; coupled with a strong correlation to the galaxies' 
\ion{H}{1} masses, they concluded that this indicates newly accreted pristine 
gas in the galaxies.  It appears that the large-scale (10 $h^{-1}$ Mpc) has 
little effect on the chemical evolution of galaxies; a galaxy's medium-scale (2 
$h^{-1}$ Mpc) environment might have much more influence on its chemical 
evolution.

\subsection{Extreme low-metallicity galaxies}

Based on observations of six extremely low-metallicity galaxies found in voids, 
\citet{Pustilnik06, Pustilnik11a, Pustilnik13} infer that there is a 
fractionally larger population of metal-poor galaxies located in voids than in 
more dense regions.  \citet{Filho15} study the environment of 140 extremely 
metal-poor galaxies and find that they preferentially reside in low-density 
environments in the local universe.  Of the 135 galaxies we analyze, twenty-one 
have extremely low gas-phase metallicity values 
($12 + \log(\text{O}/\text{H}) < 7.6$); they are highlighted in Table 
\ref{tab:lowZ}.  Of these twenty-one galaxies, only four are found in voids 
(roughly 10\% of the dwarf void population measured) and seventeen are located 
in more dense regions (about 19\% of the dwarf wall population measured).  These 
population fractions do not support the existence of a special population of 
extreme metal-poor galaxies in voids, although the statistics are very small.  
None of these galaxies share the same local environment (none are neighbors to 
each other).  In addition, Fig. \ref{fig:met1sig} shows no evidence to support a 
special population of extremely metal-poor galaxies in the voids, as extremely 
metal-poor galaxies are more prevalent in the more dense regions.

\floattable
\begin{deluxetable}{ccccccc}
\tablewidth{0pt}
\tablehead{\colhead{Index\tablenotemark{a}} & \colhead{R.A.} & \colhead{Decl.} & \colhead{Redshift} & \multicolumn{2}{c}{$12 + \log \left( \frac{\text{O}}{\text{H}} \right)$} & \colhead{Void/Wall}}
\tablecaption{Extreme low-metallicity dwarf galaxies\label{tab:lowZ}}
\startdata
268470 & \RA{13}{18}{17}{.82} & +\dec{02}{12}{59}{.83} & 0.0252 & 7.06 & $\pm$0.37 & Void \\
1422637 & \RA{14}{18}{12}{.14} & +\dec{13}{59}{33}{.98} & 0.0261 & 7.15 & $\pm$0.41 & Wall \\
839665 & \RA{08}{09}{53}{.53} & +\dec{29}{17}{04}{.82} & 0.0281 & 7.18 & $\pm$0.44 & Void \\
1168448 & \RA{11}{06}{41}{.00} & +\dec{45}{19}{09}{.28} & 0.0220 & 7.19 & $\pm$0.46 & Wall \\
1299291 & \RA{12}{17}{14}{.02} & +\dec{43}{18}{53}{.36} & 0.0233 & 7.21 & $\pm$0.42 & Wall \\
1170573 & \RA{11}{05}{39}{.42} & +\dec{46}{03}{28}{.37} & 0.0250 & 7.24 & $\pm$0.34 & Wall \\
2288717 & \RA{10}{46}{12}{.18} & +\dec{21}{31}{37}{.37} & 0.0248 & 7.27 & $\pm$0.48 & Wall \\
955643 & \RA{11}{42}{03}{.02} & +\dec{49}{21}{25}{.18} & 0.0244 & 7.28 & $\pm$0.44 & Wall \\
1344311 & \RA{12}{33}{13}{.64} & +\dec{11}{10}{28}{.46} & 0.0245 & 7.29 & $\pm$0.50 & Wall \\
1254352 & \RA{13}{29}{02}{.45} & +\dec{10}{54}{55}{.80} & 0.0237 & 7.30 & $\pm$0.44 & Wall \\
1857820 & \RA{08}{45}{00}{.34} & +\dec{27}{16}{47}{.04} & 0.0257 & 7.31 & $\pm$0.48 & Wall \\
866876 & \RA{09}{04}{57}{.96} & +\dec{41}{29}{36}{.42} & 0.0240 & 7.32 & $\pm$0.40 & Wall \\
833588 & \RA{08}{43}{10}{.71} & +\dec{43}{08}{53}{.58} & 0.0245 & 7.34 & $\pm$0.41 & Wall \\
283263 & \RA{14}{14}{12}{.88} & +\dec{01}{50}{12}{.88} & 0.0255 & 7.35 & $\pm$0.43 & Wall \\
184308 & \RA{09}{39}{09}{.38} & +\dec{00}{59}{04}{.15} & 0.0244 & 7.36 & $\pm$0.43 & Wall \\
1389829 & \RA{14}{31}{01}{.38} & +\dec{38}{04}{21}{.50} & 0.0269 & 7.41 & $\pm$0.46 & Wall \\
858951 & \RA{09}{31}{39}{.60} & +\dec{49}{49}{56}{.85} & 0.0251 & 7.46 & $\pm$0.46 & Wall \\
1270221 & \RA{13}{27}{39}{.85} & +\dec{50}{54}{09}{.69} & 0.0295 & 7.49 & $\pm$0.43 & Wall \\
431383 & \RA{08}{58}{44}{.96} & +\dec{50}{29}{58}{.98} & 0.0230 & 7.52 & $\pm$0.60 & Void \\
75442 & \RA{13}{13}{24}{.25} & +\dec{00}{15}{02}{.95} & 0.0264 & 7.55 & $\pm$0.35 & Void \\
1322765 & \RA{14}{15}{05}{.58} & +\dec{36}{22}{57}{.77} & 0.0273 & 7.57 & $\pm$0.40 & Wall\\
\enddata
\tablecomments{Details of the 21 extreme low gas-phase metallicity ($12 + \log(\text{O}/\text{H}) < 7.6$) galaxies found.  Four of these galaxies are located in voids (about 10\% of the void dwarf population measured) and seventeen are in more dense regions (about 19\% of the wall dwarf population measured); thus, there does not seem to be a special population of extreme low-metallicity galaxies in voids.  Further study of these galaxies is recommended to confirm metallicity values and identify any shared characteristics.}
\tablenotetext{a}{KIAS-VAGC galaxy index number}
\end{deluxetable}

We find that these twenty-one extremely metal-poor galaxies are redder and have 
a lower (s)SFR than the others when looking at the color (Fig. 
\ref{fig:colorZ_relation}) and (specific) star formation rate (Fig. 
\ref{fig:SFRZ_relation}) of the 135 analyzed galaxies.  The [\ion{O}{3}] 
$\lambda 4363$ auroral line is within the noise of the spectra in thirteen of 
these extremely metal-poor dwarf galaxies.  While normally such a weak detection 
of this line corresponds to a high metallicity (see Sec. \ref{sec:O3} for 
details), most of the spectra of these twenty-one galaxies have very low S/N 
overall.  As a result, it is not surprising that [\ion{O}{3}] $\lambda 4363$ is 
within the noise here.  Further study of these twenty-one galaxies is 
recommended, to confirm these low metallicity values.

\section{Conclusions}
Using spectroscopic line flux measurements of galaxies in the SDSS DR7 sample 
available through the MPA-JHU catalog, we estimate the metallicity of dwarf 
galaxies based on the Direct $T_e$ method.  From the 135 galaxies analyzed, 
there appears to be no large-scale environmental dependence of the metallicity 
of these galaxies, as the distributions of metallicity values are very similar 
for those residing in voids and those in more dense regions.  Thus, the 
large-scale ($\sim 10\text{ Mpc}$) environment does not appear to strongly 
influence the chemical evolution of dwarf galaxies.

We examine the relationship between metallicity and other physical 
characteristics of our dwarf galaxies.  In the mass-metallicity relation, our 
galaxies are at the low-mass extreme; the extreme low metallicity galaxies we 
found are scattered below this relation.  All our dwarf galaxies are at the 
upper limit in total (s)SFR, and they are on the blue end of the color spectrum.  
There is no large-scale environmental dependence of the metallicity in any of 
these categories.

No special population of extremely metal-poor galaxies is found in the voids, as 
extremely metal-deficient galaxies are found in both voids and walls.  A more 
detailed study of these twenty-one galaxies is recommended, to confirm their 
metallicity values and discover characteristics shared by the population.

Although over 800,000 galaxies in SDSS DR7 have spectroscopic observations, only 
135 are dwarf galaxies with metal line fluxes necessary to estimate gas-phase 
oxygen abundances using the Direct $T_e$ method.  Unfortunately, this was not 
enough to re-calibrate any of the more common methods used to calculate 
metallicity for use on dwarf galaxies.  Better data are required to discern the 
metallicity of a larger selection of dwarf galaxies, from which accurate 
calibrations can be developed.  These estimated ionic abundances can then be 
compared with predictions of the environmental dependence of star formation and 
metallicity from high-resolution hydrodynamic simulations.

\acknowledgements
The authors would like to thank Crystal Moorman for her help and support 
throughout this work.  We would also like to acknowledge Renyue Cen for his help 
in the interpretation of these results.  Finally, we would like to thank the 
anonymous referee for their detailed comments and critique of our work.

Support for this work was provided by NSF grant AST--1410525.

Funding for the SDSS and SDSS-II has been provided by the Alfred P. Sloan 
Foundation, the Participating Institutions, the National Science Foundation, the 
U.S. Department of Energy, the National Aeronautics and Space Administration, 
the Japanese Monbukagakusho, the Max Planck Society, and the Higher Education 
Funding Council for England.  The SDSS Web Site is \emph{http://www.sdss.org/}.

The SDSS is managed by the Astrophysical Research Consortium for the 
Participating Institutions.  The Participating Institutions are the American 
Museum of Natural History, Astrophysical Institute Potsdam, University of Basil, 
University of Cambridge, Case Western Reserve University, University of Chicago, 
Drexel University, Fermilab, the Institute for Advanced Study, the Japan 
Participation Group, Johns Hopkins University, the Joint Institute for Nuclear 
Astrophysics, the Kavli Institute for Particle Astrophysics and Cosmology, the 
Korean Scientist Group, the Chinese Academy of Sciences (LAMOST), Los Alamos 
National Laboratory, the Max-Planck-Institute for Astronomy (MPIA), the 
Max-Planck-Institute for Astrophysics (MPA), New Mexico State University, Ohio 
State University, University of Pittsburgh, University of Portsmouth, Princeton 
University, the United States Naval Observatory, and the University of 
Washington.

\bibliographystyle{apj}
\bibliography{Doug1019_sources}

\end{document}